\documentclass[12pt,draftcls,onecolumn]{IEEEtran}
\usepackage{caption}
\usepackage{subcaption}
\usepackage{booktabs}
\usepackage{setspace}
\usepackage{clrscode}
\usepackage{amsmath}
\usepackage{amssymb}
\usepackage{graphicx}
\usepackage{epsfig}
\usepackage{multirow} 
\usepackage{amsmath}  
\usepackage{relsize}
\usepackage{algorithm}
\usepackage{algorithmic}
\usepackage{amsbsy}

\usepackage{cases}
\newcommand{\E}{\mathbb{E}}

\newtheorem{Lem}{Lemma}

\newtheorem{Prop}{Proposition}

\begin{document}

\title{NOMA-based Energy-Efficient Wireless Powered Communications}

\author{Tewodros~A.~Zewde,~and~M.~Cenk~Gursoy, \textit{Senior Member, IEEE}\\
\thanks{T. A. Zewde is with the Department of Electrical Engineering and Computer Science, Wichita State University, Wichita, KS 67260 USA (email: tewodros.zewde@wichita.edu).}
\thanks{
M. C. Gursoy is with the Department of Electrical
Engineering and Computer Science, Syracuse University, Syracuse, NY, 13244
(e-mail: mcgursoy@syr.edu).}}


\maketitle

\begin{spacing}{1.8}
\begin{abstract} 
In this paper, we study the performance of non-orthogonal multiple access (NOMA) schemes in wireless powered communication networks (WPCN) focusing on the system energy efficiency (EE). We consider multiple energy harvesting user equipments (UEs) that operate based on harvest-then-transmit protocol. The uplink information transfer is carried out by using power-domain multiplexing, and the receiver decodes each UE's data in such a way that the UE with the best channel gain is decoded without interference. In order to determine optimal resource allocation strategies, we formulate optimization problems considering two models, namely half-duplex and asynchronous transmission, based on how downlink and uplink operations are coordinated. In both cases, we have concave-linear fractional problems, and hence Dinkelbach's method can be applied to obtain the globally optimal solutions. Thus, we first derive analytical expressions for the harvesting interval, and then we provide an algorithm to describe the complete procedure. Furthermore, we incorporate delay-limited sources and investigate the impact of statistical queuing constraints on the energy-efficient allocation of operating intervals. We formulate an optimization problem that maximizes the system effective-EE while UEs are applying NOMA scheme for uplink information transfer. Since the problem satisfies pseudo-concavity, we provide an iterative algorithm using bisection method to determine the unique solution.
In the numerical results, we observe that broadcasting at higher power level is more energy efficient for WPCN with uplink NOMA. Additionally, exponential decay QoS parameter has considerable impact on the optimal solution, and in the presence of strict constraints, more time is allocated for downlink interval under half-duplex operation with uplink TDMA mode.
\end{abstract}
\thispagestyle{empty}

\section{Introduction}
Wireless power transfer (WPT) is considered as a promising solution to remotely energize low-power consuming devices that might be equipped with limited-size rechargeable batteries or do not have any embedded power source at all. Additionally, WPT is more convenient to perform when wired connections are not feasible or regular battery replacement is not easily accessible, e.g., for sensors implanted in human body. In principle, WPT is carried out using electromagnetic waves or radio signals, and hence the performance depends on the wireless link characteristics and receiving circuitry design. In recent years, numerous studies in the literature have provided concrete theoretical frameworks and promising numerical results on wireless power transfer and energy harvesting (see e.g., \cite{LuX} - \cite{Ulu1} and references therein).\\
\indent As mentioned above, one advantage of WPT is to support wireless-powered communications. Each node, in these type of networks, harvests energy from either a dedicated wireless power source or ambient RF signals, and then transfers information uplink to the receiving node. Indeed, incorporating wireless-power transfer to support information transmission has a direct impact on optimal parameter values and resource allocation strategies. Hence, it is necessary to determine the optimal policies and analyze the corresponding performance characteristics \cite{ZhongC}.
\\
\indent Several studies in the literature investigated the feasibility and design of the transmission protocol, design of the receiving rectifier circuit, and downlink and uplink operation strategies of wireless-powered communication networks. The authors in \cite{Hyungisk} proposed harvest-then-transmit protocol in which an access point (AP) broadcasts wireless power to multiple users that aim to transfer information through uplink channels. In this work, the authors illustrated doubly near-far problem, i.e., sum-rate capacity maximization benefited nearby users and optimal solution encouraged to allocate more time to these users. A similar protocol was employed in \cite{Zhaof} to operate remote devices introducing average symbol error rate as a constraint while formulating an optimization problem to determine the optimal time allocation that maximizes the throughput. Further related works were presented in \cite{Hwang} and \cite{Leeh} considering multiple users that are equipped with multiple antennas, but downlink energy broadcast and uplink information transfer operations are carried out over orthogonal time intervals. Meanwhile, deploying multiple antennas at the AP or base station (BS) provides the opportunity to carry out these operations in full-duplex mode. In \cite{Ju}, the authors considered hybrid-AP that consists of two antennas to support simultaneous wireless power transfer and uplink information decoding, and incorporate the effect of self-interference as well. Each user transmitted data following the TDMA scheme, and harvested energy as long as it was not scheduled for transmission and any extra energy at the end of each block duration was stored for the next operation cycle. Similar work was presented in \cite{Kang} assuming that each user harvested energy only before it started data transmission and the harvested energy was fully utilized in each frame interval. Meanwhile, impact of statistical queuing constraints, parameterized by the quality of service (QoS) exponent, on the optimal harvesting interval for wireless information and power transfer was investigated in \cite{Teddy} where we considered half-duplex downlink-uplink operation coordination, and formulated a convex optimization problem. However, due to the difficulty in obtaining closed-form expressions, we designed an algorithm to determine the optimal solution.\\
\indent All these and related studies provide detailed analysis and interesting results considering either time/frequency-division multiplexed transmission schemes, or in general orthogonal multiple access. However, non-orthogonal multiple access (NOMA) has recently attracted much interest from both academia and industry as one of the prominent solutions for future 5G wireless networks as it enhances spectral efficiency. As discussed in the literature, NOMA is categorized into power-domain and code-domain NOMA based on how users' data multiplexing is achieved \cite{Dai}, and it can be applied to both downlink and uplink operations. In principle, power-domain NOMA utilizes superposition coding (SC) at the transmitter and successive interference cancellation (SIC) at the receiver, and this allows multiple users to transmit information on the same sub-carrier channel simultaneously. The decoding order for SIC depends on the channel characteristics of the wireless link between each transmitter-receiver pair, i.e., the main idea is that information transmitted to the receiver with the strongest wireless link is decoded without interference. In \cite{Islam}, the authors provided the basics of power-domain NOMA scheme and discussed possible solutions to address the challenges that could be experienced while applying this technique. Similarly, the authors in \cite{Datta} focused on power-domain NOMA with downlink operation, i.e., SC at the transmitter and SIC at the receivers. Another related work was presented in \cite{Lei} considering both power and channel allocation in a downlink cellular system. Meanwhile, the authors in \cite{Han} introduced and explicitly formulated the concept of power division multiple access (PDMA), and they proposed orthogonal PDMA protocol based on bit-orthogonality principle. In addition, they compared the energy efficiency of the proposed approach with conventional time/frequency division multiple access techniques.
In fact, most of the above mentioned studies analyzed the throughput to characterize and compare the performances obtained using different approaches. However, in the presence of limited power resources, efficient utilization of the available energy to transfer each bit of information is also necessary. Hence, several studies in the literature considered energy efficiency as a compelling performance metric to design optimal resource allocation strategies for future wireless networks \cite{BuzziS}. More specifically, the authors in \cite{Huu} considered heterogeneous radio access networks, and characterized the system energy efficiency in a setting in which the cloud center transferred information downlink to different types of base stations using NOMA scheme. In this work, it is argued that system energy efficiency under NOMA depends on the number of base stations in each type, and a heuristic algorithm is proposed to sequentially determine the optimal number of base stations for each type. Energy efficient resource allocation for downlink NOMA system were also presented in \cite{Huu2} and \cite{Zhang}. The authors in \cite{FF} proposed a low-complexity suboptimal algorithm for sub-channel assignment and power allocation, whereas the authors in \cite{Zhang} took into account minimum required data rate for each user. A related work was presented in \cite{SunQ} considering fading MIMO channels. Additional references on the NOMA scheme can be found in the literature e.g., in \cite{Zding} and \cite{Lsong}.\\
\indent Meanwhile, several studies have addressed the issue in regard to WPCN. In \cite{Diam}, uplink NOMA is introduced for wireless powered communications where uplink and downlink operations are carried out over non-overlapping intervals, and the authors formulated optimization problems which maximize the throughput. The authors in \cite{Yuan} studied the joint design of time allocation, downlink energy beamforming and receiver beamforming in wireless powered communication networks employing uplink NOMA. In this work, the formulated optimization problem focused on obtaining a solution that maximizes the sum rate capacity, but because of the non-convexity of the problem, an iterative algorithm was proposed. Similarly, joint optimization of base station transmit power and operating intervals for uplink NOMA in WPCNs was considered in \cite{Chin}. Yet, despite these works, the impact of NOMA on the system energy efficiency (EE) in the presence of wireless-powered users has not been investigated, to the best of our knowledge. Hence, with this motivation, we study the energy-efficient time allocation strategies for WPCN with uplink power-domain NOMA. More specifically, we consider two scenarios, namely half duplex and asynchronous transmission, based on the coordination of uplink and downlink operations, and we compare the performance gains achieved by these approaches with the conventional TDMA scheme.\\
\indent The main contributions of this paper are summarized as follows:
\begin{itemize}
\item Energy-efficient resource allocation strategies are investigated for wireless information and power transfer considering two types of uplink-downlink coordination scenarios, namely half-duplex and asynchronous transmission.
\item In both cases, we formulate optimization problems focusing on the system energy efficiency while user equipments (UEs) are allowed to transmit information-bearing signals simultaneously on the same frequency band based on non-orthogonal multiple access scheme.
\item We show that the optimization problems satisfy pseudo-concavity, and subsequently derive the necessary optimality conditions for each scenario. Due to the difficulty in obtaining analytical expressions for the optimal solution, we provide iterative algorithms using the Dinkelbach's method.
\item Using numerical results, we compare the performance gains obtained by using NOMA schemes with the conventional approaches, i.e., TDMA and OFDMA in wireless powered communication networks. For instance, we observe that downlink transmission at higher power levels improves the energy efficiency, and decoding orders can establish fairness among the users.
\item Furthermore, we consider delay-limited data sources, and address the impact of statistical queuing constraints on energy-efficient time allocation policies. In this case, we define and derive the system effective energy efficiency with downlink power transfer and uplink NOMA.
\item We formulate optimization problems that maximize the system effective energy efficiency in the presence of constraints on buffer violation probabilities at UEs. We prove the presence of unique allocation of the optimal operating intervals, and propose an algorithm based on the bisection method.
\end{itemize}
\indent The rest of this paper is organized as follows. The system model and preliminaries are discussed in Section II. An optimization problem that maximizes the system energy efficiency for half duplex operation is formulated, and an iterative algorithm is derived in Section III. In this section, optimal time allocation strategies for asynchronous scheme are also analyzed. In Section IV, delay-sensitive sources are considered and the impact of QoS constraints is studied.  Finally, numerical results are provided and conclusions are drawn in Sections V and Section VI, respectively.

\section{System Model and Preliminaries}
\subsection{System Model}
\begin{figure}
	\centering
	 \includegraphics[width=0.45\textwidth]{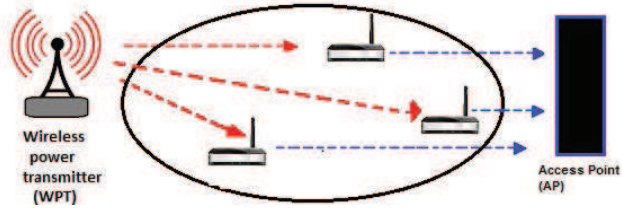}
	\caption{Network model}
	\label{Fig.ntk2}
\hrule
\end{figure}

In this paper, we consider multiple energy harvesting nodes as shown in Fig. \ref{Fig.ntk2}, which operate based on the harvest-then-transmit protocol. The wireless power transmitter (WPT) has an embedded power source, and it broadcasts a deterministic signal, denoted by $W_a$ with power $P_a=|W_a|^2$, over the downlink channel to power the nearby UEs. We employ similar assumptions as in \cite{Hyungisk} and \cite{Kang} that the $i^{th}$ user where $i\in\mathcal{S}=\{1, 2,\cdots,N\}$ fully utilizes the harvested energy to support data transmission and circuit power consumption in one cycle. Without loss of generality, we use a normalized unit for each cycle, i.e., $T=1$.\\
\indent In regard to the harvest-then-transmit protocol, UEs first harvest energy from the dedicated source (i.e., WPT), and then transmit data uplink to the access point (AP) employing the NOMA scheme\footnote{Although the WPT and AP are depicted as separate nodes in Fig. \ref{Fig.ntk2}, they can also be co-located or be the same node.}. More explicitly, we consider two scenarios, namely half-duplex operation and asynchronous transmission, based on how downlink and uplink operations are coordinated.
\\ \\
\begin{figure*}
\begin{subfigure}{0.45\textwidth}
	\centering
	 \includegraphics[width=\textwidth]{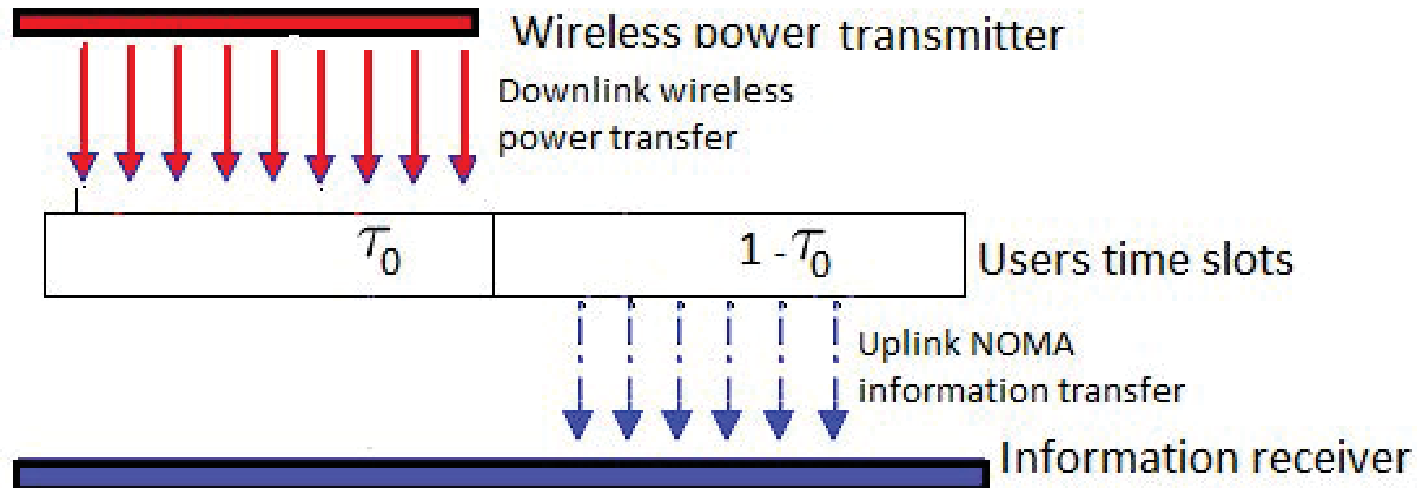}
	\caption{Half-duplex operation}
	\label{Fig.ntka}
\end{subfigure}\,\,\,\,\,\,\,\,\,\,\,\,\,\,\,\,\,
\begin{subfigure}{0.45\textwidth}
	\centering
	 \includegraphics[width=\textwidth]{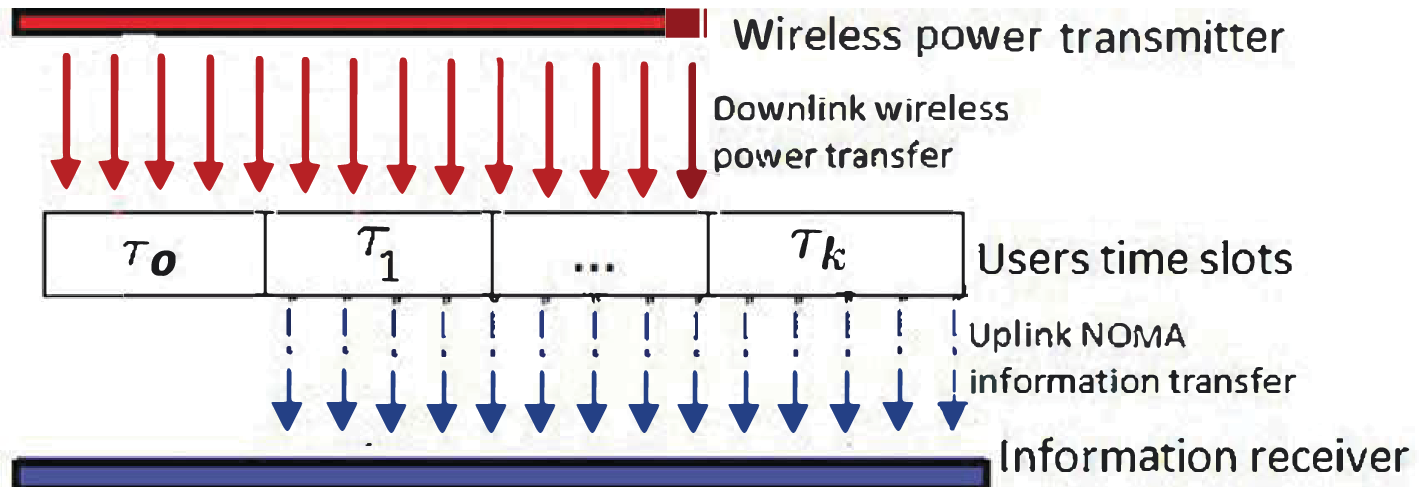}
	\caption{Asynchronous transmission}
	\label{Fig.ntkb}
\end{subfigure}
         \caption{WPCN uplink-downlink operation schemes}
	\label{Fig.ntk}
	 $\noindent\makebox[\linewidth]{\rule{18.6cm}{1pt}}$
\end{figure*}

\subsubsection{Half-duplex operation}
Here, the downlink and uplink operations are carried out over non-overlapping time intervals, i.e., all the UEs harvest energy while WPT transfers power through the downlink wireless channel over a duration $\tau_0$, and then they simultaneously transmit information-bearing signals to the AP for the rest of the period, i.e., $1-\tau_0$, as shown in Fig. \ref{Fig.ntka}. In such a case, the harvested energy at user $i\in \mathcal{S}$ in one cycle can be expressed as\footnote{Note that the formula for the harvested energy generally includes an energy harvesting efficiency factor, which we assume, without any loss of generality, to be equal to one.}
\begin{equation}
\label{eq.hv1}
E_i^{hd}=\tau_0|g_i|^2P_a \,\,\,\,\,\,\,(\text{Joules})
\end{equation}
where $\tau_0$ is the downlink energy harvesting interval which is allocated to all the users, and $g_i$ denotes the fading coefficient between the  $i^{th}$ user and WPT, and hence $|g_i|^2$ is the channel power gain.  We assume that the WPT-user links and user-AP links all experience frequency-flat fading, and uplink as well as downlink fading coefficients stay fixed in each frame duration. Thus, the received signal at the AP is expressed as
\begin{equation}
Y=\sum_{i=1}^N h_iX_i+ N_{ap}
\end{equation}
where $h_i$ is fading coefficient capturing the effect of path loss as well as small scale fading for the wireless link between user $i$ and AP, and $N_{ap} \sim \mathcal{CN}(0,1)$ is the circularly symmetric, complex Gaussian noise at the AP with unit variance. In addition, $X_i$ for $i\in\mathcal{S}$ denotes the uplink signal from UE $i$ transmitted with power $P_i$ based on power domain NOMA scheme such that $P_i>P_j$ for $|h_i|<|h_j|$. Moreover, the AP decodes the information sent from the users in the reverse order of improving channel qualities, i.e., signal from the UE with the best channel condition is decoded last without any interference from the signals transmitted by other UEs, while the signal from UE with the worst channel is decoded first in the presence of interference from all other users.\\
\indent If, without loss of generality, we assume that $|h_1|<|h_2|<\ldots<|h_N|$, then the achievable instantaneous information transfer rate of user $i$ over an uplink operation interval of $1-\tau_0$ is given as
\begin{equation}
\label{eq.R1}
R_i= (1-\tau_0)\log_2\Big(1+\frac{\gamma_i}{1+\sum_{k=i+1}^N\gamma_k}\Big) \,\,\,\,\text{bps/Hz}
\end{equation}
where $\gamma_i=|h_i|^2P_i$ is the received SNR from user $i$. Therefore, after some mathematical manipulation, the throughput or sum-rate capacity for the half-duplex scenario  becomes
\begin{equation}
\label{eq.Rsum1}
R_{sum}= \sum_{i=1}^N R_i = (1-\tau_0)\log_2\Big(1+\sum_{i=1}^N \gamma_i\Big) \,\,\,\,\text{bps/Hz}.
\end{equation}
\subsubsection{Asynchronous transmission}
In this scenario, UEs start harvesting energy at the same time, but as the name implies, they begin transmitting data signals to the AP at different time instants, as depicted in Fig. \ref{Fig.ntkb}. The advantage of this approach is that it provides an opportunity for some UEs to harvest more energy while others are scheduled for uplink information transfer. More specifically, the user with the best channel condition will be active first to send information-bearing signals to the AP. This is because the NOMA scheme encourages this UE to transmit at a lower power power level, and this can be achieved if the UE harvest energy over a shorter time interval and then use the rest of the duration to transfer information uplink. Without loss of generality, we assume that UEs are ordered according to their uplink transmission starting sequence, i.e., UE 1 begins sending data first, then user 2 and so on. Hence, the harvested energy at the $i^{th}$ UE is  given as follows:
\begin{equation}
\label{eq.hv2}
E_i^{at}=\Big(\tau_0+\sum_{j=1}^{i-1}\tau_j\Big)|g_i|^2P_a \,\,\,\,\,\,\,(\text{Joules})
\end{equation}
where $\tau_i$ is the time interval between the uplink starting points of $i^{th}$ and $(i+1)^{th}$ UEs such that
\begin{equation}
\label{eq.time}
\sum_{i=1}^N\tau_i\leq 1-\tau_0.
\end{equation}

If we assume that there are $N$ UEs, then the uplink operation time is divided into $N$ intervals, i.e., $\tau_1,\tau_2,\dots,\tau_N$, and in fact, these time slots do not necessarily have the same duration unless all UEs experience the same channel condition. During each interval, except $\tau_1$ in which only UE 1 is active for information transfer, multiple UEs send information-bearing signals to the AP, and hence the receiver applies successive interference cancellation to decode each UE's information. Thus, each UE's transmit power level depends on not only the amount of harvested energy but also the information decoding order applied at the AP. As noted above, we consider power domain NOMA scheme which encourages the UE with the best channel condition to transmit at the lowest power level as well as to be decoded without interference. Indeed, in the asynchronous operation protocol, the number of transmitting UEs increases as more time elapses. Let us denote $C|UE_{\tau_i}|$ as the total number of UEs sending information uplink during the interval $\tau_i$, and assuming $|h_1|\neq |h_2| \neq \ldots \neq|h_N|$, we have $C|UE_{\tau_i}|=i$ . Thus, during $\tau_1$ UE 1 transmits to AP, and obviously its information is decoded  without any interference. Intuitively, to comply with the NOMA scheme, the UE with the best channel condition is active first for information transfer. During $\tau_2$, UE 2, which is assumed to experience the second best channel gain, starts sending signals uplink to the AP. Hence, in this interval, we have two UEs, i.e., UE 1 and UE 2, that are actively sending information-bearing signals to the AP, and this triggers the application of successive interference cancellation to decode the information content of UE 1 and UE 2 from the received signal. The decoding order, according to the power-domain NOMA scheme, becomes in such a way that the UE with the best channel condition gets the priority to be decoded without an interference from the signal transmitted by the other UE, i.e., UE 2 is decoded first and then UE 1. In general, during the $i^{th}$ uplink operation interval, multiple UEs send information-bearing signals to the receiver, and the received signal at AP during interval $\tau_i$ is  expressed as
\begin{equation}
Y_i=\sum_{k=1}^i h_kX_k+ N_{ap}.
\end{equation}
As noted above, the receiver successively decodes each user information in the reverse order of their channel qualities. If we assume that $|h_1|\ge |h_2| \ge \ldots \ge |h_i|$, without loss of generality, then UE $i$ will be decoded first during the interval $\tau_i$ whereas UE $1$ will be the last. As a result, the achievable information rate of UE $j\in\{1,2,\cdots, i\}$ over this interval is given as
\begin{equation}
\label{eq.R2}
R_j= \tau_i\log_2\left(1+\frac{\gamma_j}{1+\sum_{k=1}^{j-1}\gamma_k}\right) \,\,\,\,\text{bps/Hz}
\end{equation}
where $|h_j|<|h_k|$ for $k=1,2,\ldots,j-1$. Note that following each incremental operating interval, $\tau_{i+1},\tau_{i+2},\dots,\tau_N$, one or more UEs join the uplink operation, and hence the information decoding order at the receiving end is modified accordingly.
The total throughput becomes the sum of the sum-rate capacity of the system over each interval until the end of $\tau_N$.
After some mathematical manipulations, we have
\begin{equation}
\label{eq.Rsum2}
R_{sum}=\sum_{i=1}^k \tau_i \log_2\left(1+\sum_{j=1}^i\gamma_j\right).
\end{equation}
\subsection{Energy Efficiency}
In wireless communication systems, user energy efficiency (EE) can be quantitatively measured by the bits of information reliably transferred to a receiver per unit consumed energy at the transmitter. In the presence of multiple users, it is also relevant and meaningful to consider system EE. Additionally, this enables the allocation of the resources in such a way that the overall energy usage becomes more efficient. With this motivation, we consider system energy efficiency which is defined as
\begin{equation}
\eta=\frac{\text{Throughput}}{\text{Total consumed energy}}\,\,\,\,\,\,\,\,\,\,\,\,\,(\text{bits/Joule}).
\end{equation}
The sum-rate capacity is maximized when each source transmits at its peak power level. However, this might not be the optimal strategy when energy efficiency is considered.\\
\indent In practice, energy is consumed to power up data processing circuitry and send the signal to the target destination. Let $P_{c_D}$ denote the circuit power consumption at WPT during downlink operation, and assume that it is independent of the transmitted power level for $P_a>0$. However, if no wireless power is transferred, there is no consumption, i.e., $P_a=0$ and $P_{c_D}=0$. Hence, the total energy consumption during the entire downlink-uplink operation of a given cycle becomes
\begin{equation}
\label{Eq.Etot}
E_{tot}=\left\{
\begin{array}{ll}
\tau_0P_{DT}+(1-\tau_0)P_{c_U}&\hbox{Half-duplex}\\~\\
\Big(\tau_0+\sum_{i=1}^{N-1}\tau_i\Big)P_{DT}+P_{c_U}\sum_{i=1}^N\tau_i&\hbox{Asynchronous}
\end{array}
\right.
\end{equation}
where $P_{DT}=P_{c_D}+P_a$, and $P_{c_U}$ is the power consumption at the receiver for decoding information during uplink operation.

\subsection{Delay-Sensitive Sources and Effective Capacity}

In this subsection, we address how to determine the throughput if data to be sent from the UEs are delay-sensitive. In particular, we introduce effective capacity as the throughput metric in the presence of statistical queueing constraints.
We assume that, while harvesting energy, each user stores received data packets generated by a delay-sensitive source that requires certain statistical QoS guarantees described by the QoS exponent $\theta$. More specifically, the tail distribution of the buffer length is required to have an exponential decay with rate controlled by the exponent $\theta$, and hence the buffer violation or overflow probability is described as
\begin{equation}
\label{Eq.4}
\Pr\Big\{Q_i\geq Q_{max}\Big\}\approx e^{-\theta_i Q_{max}}
\end{equation}
where $Q_i$ denotes the stationary queue length in the $i^{th}$ user buffer, and $Q_{max}$ is the buffer overflow threshold. This buffer constraint limits the arrival rates that can be supported by the wireless link.

Instantaneous channel capacity provides the maximum achievable data rate at which information can be transmitted through the wireless medium. However, arrival rates at which data packets are received from the source are further limited by the buffering requirements such as the statistical queuing constraints described in (\ref{Eq.4}). Let $r_i[n]$ and $R_i[n]$ denote $i^{th}$ user random arrival rate and instantaneous transmission (or equivalently service) rate, respectively, in the $n^{th}$ time slot. The corresponding asymptotic logarithmic moment generating functions (LMGF) $\Lambda_A$ and $\Lambda_C$ of the arrival and service processes, are given as follows \cite{Chang}:
\begin{equation}
\begin{split}
\label{Eq.LMGF}
\Lambda_A&=\lim_{t\rightarrow\infty} \frac{\log\Big(\E\big\{e^{\theta \sum_{n=1}^tr_i[n]}\big\}\Big)}{t}\\
\Lambda_C&=\lim_{t\rightarrow\infty} \frac{\log\Big(\E\big\{e^{\theta \sum_{n=1}^tR_i[n]}\big\}\Big)}{t}.
\end{split}
\end{equation}
Having the buffer overflow probability to decay exponentially with rate $\theta$ as in (\ref{Eq.4}) requires \cite{Chang}
\begin{gather}
\Lambda_A(\theta)+\Lambda_C(-\theta)=0. \label{eq:lmgf-equality}
\end{gather}
In this paper, we consider constant data arrival rates, i.e., we assume $r_i[n]=r$ $\forall n$, which leads to $\Lambda_A = \theta r$. Now, by solving (\ref{eq:lmgf-equality}), we obtain the maximum constant arrival rate, also termed as effective capacity, for a given certain QoS exponent $\theta$ as follows \cite{dapeng}:
\begin{equation}
\label{eq.effCC}
C_i^e(\theta_i)=-\lim_{t\rightarrow \infty}\frac{1}{t\theta_i}\log\bigg(\E\left\{e^{-\theta_i \sum_{n=1}^t R_i[n]}\right\}\bigg).
\end{equation}
Assuming block fading with block duration $T$, this can be further simplified as\\
\begin{equation}
\label{eq.effC}
C_i^e(\theta_i)=-\frac{1}{T\theta_i}\log\bigg(\E\big\{e^{-\theta_iT R_i[n]}\big\}\bigg).
\end{equation}
As the buffer constraint is relaxed, effective capacity approaches ergodic capacity, i.e., $\lim_{\theta_i\rightarrow 0} C_i^e=\E\{R_i\}$, whereas for increasingly strict constraints i.e., as $\theta_i\rightarrow \infty$, effective capacity converges to the delay-limited capacity with zero outage.

\section{energy-efficient time allocation strategies}

In this section, we analyze time allocation strategies for energy harvesting and data transmission phases, considering both half-duplex and asynchronous operations. No delay or buffer constraints are imposed initially. Delay-sensitive sources and statistical QoS constraints are introduced into the analysis in Section IV.

\subsection{Optimal harvesting interval with half-duplex operation}
Assuming that the harvesting interval depends on the fading state realizations, the uplink transmitted signal power level from the $i^{th}$ user becomes
\begin{equation}
\begin{split}
\label{eq.Pi1}
P_i=\xi_i|g_i|^2P_a\frac{\tau_0}{1-\tau_0}.
\end{split}
\end{equation}
Note that $\xi_i$ denotes the fraction of harvested energy utilized for data transmission while the rest, i.e., the fraction of $1-\xi_i$, is consumed by the circuit to carry out the process. Then, substituting (\ref{eq.Pi1}) into (\ref{eq.R1}) and simplifying the expression, we get
\begin{equation}
\label{eq.R1tau}
\mathcal{R}_i(\tau_0)=(1-\tau_0)\log_2\Bigg(1+\frac{\alpha_i\tau_0}{1-\tau_0+\sum_{j=i+1}^N\alpha_j\tau_0}\Bigg)
\end{equation}
where $\alpha_i=\xi_i|g_i|^2|h_i|^2P_a$. We first have the following characterization.
\begin{Lem}
The individual achievable rate of a wireless-powered UE operating in half-duplex mode with uplink NOMA strategy is concave in the harvesting interval $\tau_0$.
\end{Lem}
\emph{Proof}: The expression given in (\ref{eq.R1tau}) can be re-written as
\begin{equation}
\label{eq.R1taunew}
\mathcal{R}_i(\tau_0)=(1-\tau_0)\log_2\Bigg(1+\frac{\alpha_i\tau_0}{1+\omega_i\tau_0}\Bigg)
\end{equation}
where $\omega_i=-1+\sum_{j=i+1}^N\alpha_j$. Hence, applying the second order derivative criteria to (\ref{eq.R1taunew}), we have
\begin{equation}
\frac{\partial^2 R_i(\tau_0)}{\partial \tau_0^2}\!=\!-\alpha_i\!\left[\!\frac{\alpha_i((2\omega_i+1)\tau_0+1)+2(\omega_i+1)(\omega_i\tau_0+1)}{(\omega_i\tau_0)^2(\alpha_i\tau_0+\omega_i\tau_0+1)^2}\right]
\end{equation}
Knowing $\omega_i\geq-1$ and $0<\tau_0<1$, it is obvious that $(2\omega_i+1)\tau_0\leq 1$ if $\omega_i<0$, otherwise $(2\omega_i+1)\tau_0> 1$. This guarantees $\frac{\partial^2 R_i(\tau_0)}{\partial \tau_0^2}<0$ for any $\alpha_i\neq 0$, and hence $R_i$ is a concave function. This completes the proof. $\hfill\blacksquare$\\	
\indent Note that the total throughput, i.e., the sum of individual achievable data rates, is given by
\begin{equation}
\begin{split}
\mathcal{R}_{sum}(\tau_0)=&\sum_{i=1}^N\mathcal{R}_i(\tau_0)\\
=&(1-\tau_0)\log_2\Bigg(1+\sum_{i=1}^N \alpha_i\frac{\tau_0}{1-\tau_0}\Bigg).
\end{split}
\end{equation}
Therefore, the system energy efficiency (EE), which measures the numbers of bits of information reliably transmitted to the AP per consumed unit energy, is given as
\begin{equation}
\label{Eq.EE1}
\eta_{HD}(\tau_0)=\frac{(1-\tau_0)\log_2\Bigg(1+\sum_{i=1}^N \alpha_i\frac{\tau_0}{1-\tau_0}\Bigg)}{\tau_0\big(P_{c_D}+P_a-P_{c_U}\big)+P_{c_U}}.
\end{equation}
Note that each UE's circuit power consumption is supported by the harvested energy, and hence it is not necessary to consider these explicitly while defining the total energy consumption of the system.
\begin{Prop}
The system EE of a wireless-powered communication network given in (\ref{Eq.EE1}) is a pseudo-concave function of the harvesting interval $\tau_0$.
\end{Prop}
\emph{Proof}: Since the system EE in (\ref{Eq.EE1}) is a fractional function, it will satisfy pseudo-concavity according to Proposition 2.9 stated in \cite{Alessio} if the numerator is concave and denominator is convex. In this case, the denominator is an affine function, and hence we only need to show that the throughput, i.e., the numerator, is concave with respect to $\tau_0$. Using the fact that concavity is preserved under summation and $R_i(\tau_0)$ is a concave function of $\tau_0$ based on Lemma 1, we conclude that $\mathcal{R}_{sum}$, i.e., the throughput, is concave as well. This completes the proof.$\hfill\blacksquare$\\
\indent Proposition 1 guarantees that there is a unique optimal time allocation strategy that maximizes the system EE such that the harvesting interval is within the feasible set. 
In order to obtain the optimal time allocation for downlink and uplink operations that maximizes the system energy efficiency, we formulate the following optimization problem:
\begin{subequations}
\begin{align}
\label{eq.obj1}
(\text{PR:1})\,\,\,\,\,\,\,\,\,\max_{\tau_0}\,\,\,& \eta_{HD}(\tau_0)\\
\label{eq.const1}
 \text{subject to}\,\,\,&\tau_0(1-\tau_0)\ge0.
\end{align}
\end{subequations}
\indent The constraint in (\ref{eq.const1}) dictates that the optimizing parameter $\tau_0$ is always within the feasible set, i.e., $0 \le \tau_0 \le 1$. Since this constraint is convex, and the objective function is the ratio of a concave function over an affine function, it is obvious that (PR:1) is a concave-linear fractional programming (CLFP) problem. As noted in \cite{Alessio}, Dinkelbach's method can be used to solve concave-convex and concave-linear fractional programming problems, and we employ this method to identify the optimal solution. Thus, (PR:1) can be equivalently expressed as
\begin{subequations}
\begin{align}
\label{eq.obj11}
\min_{\lambda}\,\,\,\bigg\{\max_{\tau_0}\,\,&\,\,\mathcal{L}(\tau_0,\lambda)\bigg\}\\
 \text{subject to}\,\,\,&\,\,(\ref{eq.const1})
\end{align}
\end{subequations}
where
$\mathcal{L}(\tau_0,\lambda)\!=\!(\!1\!-\!\tau_0\!)\log_2\!\Big(\!1\!+\!\frac{\alpha_T\tau_0}{1-\tau_0}\!\Big)-\lambda\Big(\tau_0 P_{\Delta}\!+\!P_{c_D}\!\Big)$, $\alpha_T=\sum_{i=1}^N \alpha_i$ and $P_\Delta=P_{c_D}+P_a-P_{c_U}$. Since the achievable rate is concave function on downlink operating interval $\tau_0$ and the total energy consumption, $\tau_0P_\Delta+P_{c_D}$, is linear, it is obvious that the Lagrangian $\mathcal{L}(\tau_0)$ is concave with respect to the harvesting interval $\tau_0$. We know that the constraint is convex, and hence, the inner optimization problem is a concave maximization or equivalently a convex minimization problem. In such a case, Karush-Kuhn-Tucker (KKT) conditions, i.e.,
\begin{subequations}
\begin{align}
\label{Eq.L1a}
\frac{\partial \mathcal{L}}{\partial \tau_0}\bigg|_{\tau_0=\tau_0^\ast}=&0\\
\label{Eq.L1b}
\mu^\ast\Big(1-\tau_0^\ast\Big)=0,\,\,\,\,\,\,\,\,\,\kappa^\ast\tau_0^\ast=&0,
\end{align}
\end{subequations}
are necessary and sufficient for the global optimality of the solution given the dual parameter $\lambda$. From the characteristics of the EE curve, when $\tau_0^\ast=0$ or $\tau_0^\ast=1$, we have $\eta_{HD}(\tau_0^\ast)=0$ for $P_{c_D}\neq 0$. But, this cannot be the optimal value, which implies $0<\tau_0^\ast<1$. As a result, $\mu^\ast=0$ and $\kappa^\ast=0$ in order to satisfy the complementary slackness conditions given in (\ref{Eq.L1b}). Taking these into account, and applying the first order optimality criteria given in (\ref{Eq.L1a}), we obtain
\begin{equation}
\frac{\alpha_T}{1-\tau_0+\alpha_T\tau_0}-\ln\left(1+\alpha_T\frac{\tau_0}{1-\tau_0}\right)\!-\!\ln(2)\lambda P_\Delta=0
\end{equation}
which leads to
\begin{equation}
z\ln(z)+\Omega z=\alpha'
\end{equation}
or equivalently
\begin{equation}
\label{eq.z1}
e^{\ln(z e^{\Omega})}\ln(z e^{\Omega})=\alpha' e^{\Omega}
\end{equation}
where $z=1+\alpha_T\frac{\tau_0}{1-\tau_0}$, $\Omega=\ln(2)\lambda P_\Delta-1$, and $\alpha'=\alpha_T-1$. Mathematically, (\ref{eq.z1}) has the form of $Xe^X=Y$ whose solution is given by the Lambert function, i.e., $X=\mathcal{W}(Y)$ for $Y\geq -\frac{1}{e}$. Thus, the solution to (\ref{eq.z1}) can be analytically expressed as
\begin{equation}
\label{eq.zstr0}
z^\ast=\bold{e}^{\left[\mathcal{W}(\alpha' .\bold{e}^{\Omega})-\Omega\right]}.
\end{equation}
Based on the definition of $z$ and substituting the expression given in (\ref{eq.zstr0}), and using $\alpha_T=\alpha'-1$, the solution for the optimal harvesting time can be expressed as
\begin{equation}
\begin{split}
\tau_0=&\frac{z-1}{\alpha_T+z-1}\\
=&\frac{e^{\mathcal{W}(\alpha' .\bold{e}^{\Omega})}-e^{\Omega}}{\alpha'e^{\Omega}+e^{\mathcal{W}(\alpha' .\bold{e}^{\Omega})}}.
\end{split}
\end{equation}
Applying the property $e^{\mathcal{W}(x)}=\frac{x}{\mathcal{W}(x)}$ to the above equation, and after several manipulations, we get
\begin{equation}
\label{eq.tau_sol}
\tau_0^\ast(\lambda)=\frac{\alpha'-\mathcal{W}\Big(\alpha' .\bold{e}^{(\lambda P'-1)}\Big)}{\alpha'\Big[1+\mathcal{W}\Big(\alpha' .\bold{e}^{(\lambda P'-1)}\Big)\Big]}
\end{equation}
where $P'=\ln(2)P_\Delta$. The optimal harvesting time $\tau_0^\ast$ is a function of the parameter $\lambda$, and this parameter is iteratively updated until the optimal solution satisfies $\mathcal{R}_{sum}(\tau^\ast)-\lambda^\ast E_{tot}(\tau^\ast)=0$. We provide the complete procedure below in Algorithm \ref{alg00}.
\begin{algorithm}                      
	\caption{EE maximization using Dinkelbach's algorithm}
	\label{alg00}                           
	\begin{algorithmic}[1]                 
		\STATE Given: $\epsilon$, $\lambda_0$
		\STATE $n\leftarrow 0$
		\REPEAT 	
		\STATE {Determine $\tau_0^\ast$ using (\ref{eq.tau_sol})}
		\STATE {$\mathcal{F}(\lambda_n,\tau_0^\ast)=\mathcal{R}_{sum}(\tau_0^\ast)-\lambda_n E_{tot}(\tau_0^\ast)$}
		\STATE {$\lambda_{n+1}=\frac{\mathcal{R}_{sum}(\tau_0^\ast)}{E_{tot}(\tau_0^\ast)}$}
		\STATE {$n\leftarrow n+1$}		
		\UNTIL {$|\mathcal{F}(\lambda_n,\tau_0^\ast)|<\epsilon$}	
		\STATE Set $\tau_0^*=\tau_0^n$.
	\end{algorithmic}
\end{algorithm}
\subsection{Energy-efficient intervals with asynchronous transmission}
\subsubsection{Overlapping uplink operation}
\indent As noted above in Section II, asynchronous transmission is defined in such a way that UEs do not necessarily begin sending information-bearing signals to the AP at the same time in each downlink-uplink operation cycle. However, if a UE has started transmission, it stays active until the end of the cycle. Assuming that UEs are ordered according to their uplink starting point as mentioned earlier, the transmitted signal power level from UE $i$ is given as
\begin{equation}
\begin{split}
\label{eq.p2}
P_i=&\xi_i\frac{E_i^{at}}{\sum_{k=i}^N\tau_k}\\
=&\xi_i|g_i|^2P_a\left[\frac{\tau_0+\sum_{j=1}^{i-1} \tau_j}{\sum_{k=i}^N\tau_k}\right]
\end{split}
\end{equation}
where $E_i^{at}$ denotes the harvested energy by UE $i$ in the asynchronous scheme, and $\xi_i$ is the fraction of harvested energy utilized for data transmission while the rest, i.e., the fraction of $1-\xi_i$, is consumed by the circuit to carry out the process.\\
\indent Then, we substitute (\ref{eq.p2}) into (\ref{eq.R2}), and derive the expression for the sum-rate capacity (within the interval of duration $\tau_i$ in which $i$ users are transmitting) as a function of the operating intervals as
\begin{equation}
\mathcal{R}_{sum}^i(\tau_0,\boldsymbol{\tau}_N)=\tau_i\log_2\Bigg(1+\sum_{l=1}^ib_l \frac{\tau_0+\sum_{j=1}^{l-1} \tau_j}{\sum_{k=l}^N\tau_k}\Bigg)
\end{equation}
which leads to
\begin{equation}
\label{eq.Rsmi}
\mathcal{R}_{sum}^i(\tau_0,\boldsymbol{\tau}_N)=\tau_i\log_2\Bigg(a_i+ \sum_{l=1}^i\frac{b_l}{\sum_{k=l}^N\tau_k}\Bigg)
\end{equation}
where $b_i=\xi_i|g_i|^2|h_i|^2P_a$, $a_i=1-\sum_{l=1}^i b_l$ and $\boldsymbol{\tau}_N=[\tau_1,\tau_2,\cdots,\tau_N]$.
\begin{Lem} \label{lemma:lemma2}
The achievable sum-rate capacity for the two-user setting during the transmission interval of duration $\tau_i$ is jointly concave over the operating intervals $(\tau_0, \tau_1,\tau_2)$.
\end{Lem}

\emph{Proof}: See Appendix \ref{app:lemma2proof}.

From Lemma 2, we conclude that the system throughput is a jointly concave function since concavity is preserved under summation. Note that the total sum-rate capacity is given as
\begin{equation}
\mathcal{R}_{tot}(\boldsymbol{\tau}_N)=\sum_{i=1}^2\tau_i\log_2\Bigg(a_i+ \sum_{l=1}^i\frac{b_l}{\sum_{k=l}^2\tau_k}\Bigg).
\end{equation}
Then, the system energy efficiency (EE) for asynchronous transmission scenario becomes
\begin{equation}
\label{Eq.EE2}
\eta_{AT}(\tau_0,\tau_1,\tau_2)\!=\!\frac{\tau_1\log_2\Big(\!a_1\!+\!\frac{b_1}{\tau_1+\tau_2}\!\Big)\!+\!\tau_2\log_2\Big(\!a_2\!+\!\frac{b_1}{\tau_1+\tau_2}\!+\!\frac{b_2}{\tau_2}\Big)\!}{P_{DT}\big(\tau_0+\tau_1\big)+\big(\tau_1P_{c_U}^1+\tau_2P_{c_U}^2\big)}
\end{equation}
where $P_{c_U}^i$ denotes the total uplink circuit power consumption during the interval $\tau_i$.\\
\indent Since the throughput is proved to be a concave function, and the total consumed power is an affine function of the operation intervals, the system EE given in (\ref{Eq.EE2}) satisfies the criteria for pseudo-concavity based on Proposition 2.9 stated in \cite{Alessio}. Unlike the previous scenario where we had only one parameter to adjust, i.e., $\tau_0$, to achieve maximum energy efficiency, now there are $3$ optimizing parameters, i.e., $\tau_0,\tau_1,\tau_2$, and hence obtaining the optimal time allocation strategy which maximizes the EE is a more challenging task. Thus, we formulate the following optimization problem:
\begin{subequations}
\begin{align}
(\text{PR:3A})\,\,\,\,\,\,\,\,\,\max_{\tau_0,\boldsymbol{\tau}_2}\,\,\,& \eta_{AT} \label{eq.objPR3A}\\
\label{eq.const2a1}
 \text{subject to}\,\,\,&\sum_{i=1}^2\tau_i\leq 1- \tau_0\\
 \label{eq.const2a2}
         & \tau_i\geq 0,\,\,\,\,\,\,\,\,\,\,\,i\in\{0,1,2\}.
\end{align}
\end{subequations}
As noted above, the objective function in (\ref{eq.objPR3A}) is pseudo-concave since the total achievable sum-rate capacity is jointly concave with respect to operating intervals $(\tau_0,\tau_1,\tau_2)$. In addition, the constraints (\ref{eq.const2a1}) and (\ref{eq.const2a2}), which define the feasible operating intervals, are convex. Thus, the optimization problem (PR:3) is also a concave-linear fractional programming problem, and hence it can be easily solved using Dinkelbach's algorithm following a similar procedure as in the earlier scenario, but we skip the details for brevity.\\
\subsubsection{Non-overlapping uplink operation}
In this subsection, we consider a special scenario where energy harvesting by a user can still occur concurrently with the data transmission of other users, but the uplink data transmission among users follows time-division multiple access instead of allowing the activated user to use the channel until the end of the block duration. Hence, data transmissions by the users occur over non-overlapping time intervals. In such a case, the system energy efficiency expression for $N$ users is given by
\begin{equation}
\label{Eq.EE3}
\eta_{AT}(\tau_0,\boldsymbol{\tau}_N)=\frac{\sum_{i=1}^N\tau_i\log_2\Bigg(1+ b_i \frac{\sum_{k=0}^{i-1}\tau_k}{\tau_i}\Bigg)}{P_{DT}\big(\tau_0+\sum_{i=1}^{N-1}\tau_i\big)+P_{c_U}\sum_{i=1}^N\tau_i}
\end{equation}
where $P_{c_U}$ denotes the circuit power consumption of a UE assuming that each UE consumes the same amount. Therefore, the optimization problem is reformulated as follows:
\begin{subequations}
	\begin{align}
	 (\text{PR:3B})\,\,\,\,\,\,\,\,\,\max_{\tau_0,\boldsymbol{\tau}_N}\,\,\,& \eta_{AT}\\
	\label{eq.const3a1}
	\text{subject to}\,\,\,&\sum_{i=1}^N\tau_i\leq 1- \tau_0\\
	\label{eq.const3a2}
	& \mathcal{R}_i(\tau_0,\boldsymbol{\tau}_i)\geq R^i_{min},\,\,\,\,\,\forall i\in\mathcal{S}
	\end{align}
\end{subequations}
where
\begin{equation} \mathcal{R}_i(\tau_0,\boldsymbol{\tau}_i)=\tau_i \log_2\Bigg(1+b_i\frac{\tau_0+\sum_{j=1}^{i-1} \tau_j}{\tau_i}\Bigg)
\end{equation}
and $\boldsymbol{\tau}_i=[\tau_1,\cdots,\tau_i]$. The additional constraint given in (\ref{eq.const3a2}) is introduced in order to guarantee that each user's rate is above a certain minimum level when non-overlapping time slots are being allocated to the users. From Lemma 2 in \cite{Kang}, we know that $\mathcal{R}_i$ is a jointly concave function of downlink and uplink time intervals, i.e., $\tau_0$ and $\boldsymbol{\tau}_i$. Thus, the above optimization problem is still a concave-linear fractional programming problem, and (PR:3B) can be equivalently expressed as
\begin{subequations}
	\begin{align}
	\label{eq.obj111}
	 \min_{\lambda,\mu}\,\,\,&\bigg\{\max_{\tau_0,\boldsymbol{\tau}_i}\,\,\,\,\mathcal{G}(\tau_0,\boldsymbol{\tau}_i)\bigg\}\\
	\text{subject to}\,\,\,&\,\,(\ref{eq.const3a1})\,\,\,\,\text{and}\,\,\,\,(\ref{eq.const3a2})
	\end{align}
\end{subequations}
where
$\mathcal{G}(\tau_0,\boldsymbol{\tau}_i)\!=\!\!\sum_{i=1}^N\tau_i\log_2\!\Big(\!1\!+b_i\!\frac{\!\tau_0+\sum_{j=1}^{i-1}\! \tau_j\!}{\tau_i}\Big)\!-\lambda\Big(\!P_{DT}\big(\tau_0\!+\!\sum_{i=1}^{N-1}\tau_i\big)+P_{c_U}\sum_{i=1}^N\tau_i\!\Big)$, and $P_\Delta=P_{DT}-P_{c_U}$. Given $\lambda$, the objective function for the inner maximization problem $\mathcal{G}(\tau_0)$ is a concave function while the constraints are convex, and hence Karush-Kuhn-Tucker (KKT) conditions, i.e.,
\begin{subequations}
	\begin{align}
	\label{Eq.L1aaa}
	&\frac{\partial \mathcal{L}}{\partial \tau_0}\bigg|_{\tau_0=\tau_0^\ast}=0,\,\,\,\,\,\,\,\,\,\,\,\,\frac{\partial \mathcal{L}}{\partial \tau_i}\bigg|_{\tau_i=\tau_i^\ast}=0\\
	\label{Eq.L1bb}
	 &\kappa_i^\ast\Big( \mathcal{R}_i(\tau_0,\boldsymbol{\tau}_i)-R^i_{min}\Big)=0,\,\,\,\forall i\in\mathcal{S}\\
	 &\zeta^\ast\left(\sum_{i=0}^N\tau_i-1\right)=0,
	\end{align}
\end{subequations}
are necessary and sufficient for global optimality where the Lagrangian is given as
\begin{equation}
\mathcal{L}=\mathcal{G}(\tau_0,\boldsymbol{\tau}_i)+\zeta\Big(\sum_{i=0}^N\tau_i-1\Big)+\sum_{i=1}^N\kappa_i\Big(R^i_{min}- \mathcal{R}_i(\tau_0,\boldsymbol{\tau}_i)\Big).
\end{equation}
{\small
\begin{figure*}
\small
	\begin{subequations}
		\label{eq.nov}
		\begin{align}
		\label{eq.nova}
		\frac{\partial \mathcal{L}}{\partial \tau_0}&=\sum_{k=1}^N \frac{(1-\kappa_k)b_k}{\ln(2)\!\Big(\!1\!+b_k\!\frac{\!\tau_0+\sum_{j=1}^{i-1}\! \tau_j\!}{\tau_k}\!\Big)}-\lambda P_{DT}+\zeta=0\\
		\label{eq.novb}
		\frac{\partial \mathcal{L}}{\partial \tau_i}&=\sum_{k> i}^N \frac{(1-\kappa_k)b_k}{\ln(2)\!\Big(\!1\!+b_k\!\frac{\!\tau_0+\sum_{j=1}^{i-1}\! \tau_j\!}{\tau_k}\!\Big)}+(1-\kappa_i)\log_2\!\Big(\!1\!+b_i\!\frac{\!\tau_0+\sum_{j=1}^{i-1}\! \tau_j\!}{\tau_i}\!\Big)-\frac{(1-\kappa_i)b_i\!\frac{\!\tau_0+\sum_{j=1}^{i-1}\! \tau_j\!}{\tau_i}}{\ln(2)\!\Big(\!1\!+b_i\!\frac{\!\tau_0+\sum_{j=1}^{i-1}\! \tau_j\!}{\tau_i}\!\Big)}-\lambda (P_{DT}+P_{c_U})+\zeta=0\\
		\label{eq.novc}
		\frac{\partial \mathcal{L}}{\partial \tau_N}&=(1-\kappa_N)\log_2\Big(1+b_i\frac{\tau_0+\sum_{j=1}^{N-1} \tau_j}{\tau_N}\Big)-\frac{(1-\kappa_N)b_N\frac{\tau_0+\sum_{j=1}^{N-1}\tau_j}{\tau_N}}{\ln(2)\Big(1+b_N\frac{\tau_0+\sum_{j=1}^{N-1} \tau_j}{\tau_N}\Big)}-\lambda P_{c_U}+\zeta=0
		\end{align}
	\end{subequations}
	 $\noindent\makebox[\linewidth]{\rule{18.5cm}{0.5pt}}$
\end{figure*}}
\normalsize
\\
Applying the first order optimality criterion in (\ref{Eq.L1aaa}), we obtain the optimality conditions in (\ref{eq.nov}) given at the top of this page.
Taking the difference of (\ref{eq.nova}) and (\ref{eq.novb}), we have
\begin{equation}
-\sum_{k=1}^i \frac{(1-\kappa_k)b_k}{1+b_kz_k}+(1-k_i)\mathcal{Z}_i(z_i)+\lambda P_{c_U}\ln(2)=0
\end{equation}
which leads to
\begin{equation}
\mathcal{Z}_i(z_i) -\frac{b_i}{1+b_iz_i}=\frac{\lambda P_{c_U} \ln(2)}{1-\kappa_i}+\sum_{j=1}^{i-1}\frac{b_j}{1+b_jz_j}
\end{equation}
where $\mathcal{Z}_i(z_i)=\ln(1+b_iz_i)-\frac{b_iz_i}{1+b_iz_i}$ and $z_k=\frac{\tau_0+\sum_{j=1}^{k-1}\tau_j}{\tau_k}$. Similarly, from (\ref{eq.novc}), we have
\begin{equation}
\mathcal{Z}_N(z_N)-\frac{b_N}{1+b_Nz_N}=\frac{\lambda (P_{DT}+P_{c_U})\ln(2)}{1-\kappa_N}+\sum_{k=1}^{N-1} \frac{b_k}{1+b_kz_k}.
\end{equation}
Applying a similar approach as in \cite{Kang}, the optimal time allocations are given as
\begin{subequations}
\begin{align}
\label{eq.taustrN}
\tau_N^\ast&=\frac{1}{1+z_N}\\
\label{eq.taustri}
\tau_i^\ast&=\frac{1-\sum_{j=i+1}^N\tau_j^\ast}{1+z_i},\,\,\,\,\,\text{for }i=1,2,\dots,N-1
\end{align}
\end{subequations}
where
\begin{equation}
\label{eq.zstr}
z_i=\frac{1}{b_i}\left[e^{\mathcal{W}\Big(\frac{b_i-1}{e^{\phi_i+1}}\Big)+\phi_i+1}-1\right],\,\,\,\,i\in\mathcal{S}
\end{equation}
with
\begin{subequations}
\begin{align}
\label{Eq.phi_i}
\phi_i&=\frac{\lambda P_{c_U} \ln(2)}{1-\kappa_i}+\sum_{j=1}^{i-1}\frac{b_j}{1+b_jz_j}, \,\,\,\,\,i=1,2,\dots,N-1\\
\label{Eq.phi_N}
\phi_N&=\frac{\lambda \Big(P_{DT}+P_{c_U}\Big)\ln(2)}{1-\kappa_N}+\sum_{k=1}^{N-1} \frac{b_k}{1+b_kz_k}.
\end{align}
\end{subequations}
Based on (47), we can see that, for instance, the $i^{th}$ uplink operating interval $\tau_i$ is dependent on the time intervals $\tau_N,\tau_{N-1},\dots,\tau_{i-1}$, and hence we should compute these intervals in order to determine $\tau_i$. This implies that the operating intervals need to computed sequentially, i.e., first $\tau_N$, then $\tau_{N-1}$, and so on until $\tau_1$. However, in order to compute $\tau_i$, $z_i$ has to be known. It is obvious that the parameter $z_i$ needs be computed sequentially, but according to (48) and (49), its order is different from $\tau_i$s, i.e., first $z_1$, second $z_2$, and so on until $z_N$. Since $z_i$ depends on the parameter $\phi$, which is explicitly defined in (49), we need to determine $\phi_1$ first and then substitute its solution in (48) to compute $z_1$. This procedure continues iteratively until we compute all the values of $z_i$s.\\
\indent Hence, the optimal time interval, $\tau_0$, becomes $\tau_0^\ast=1-\sum_{i=1}^N\tau_i^\ast$. From the above expressions, we observe that the energy-efficient time allocation depends on the minimum data rate constraint. For instance, if this constraint is inactive for all UEs, then $\kappa_i=0$ $\forall i$ due to complementary slackness conditions. However, if it is active for any UE, then the corresponding optimal solution will be changed in such a way that the constraint is satisfied while maximizing the system energy efficiency. Therefore, we first determine the best solution assuming all the rate constraints are satisfied with inequality, i.e., $\kappa_i=0$ $\forall i\in\mathcal{S}$ and then check if the optimal solution satisfies the rate constraint for each UE. For any constraint violation, the optimal time allocation policy will be updated taking into account all of the active constraints, and the detailed procedure is provided in Algorithm \ref{Algo2} on the next page.
\begin{algorithm}                      
	\caption{Energy-efficient time allocation for non-overlapping scheme}
	\label{Algo2}                           
	\begin{algorithmic}[1]                 
		\STATE Given: $\epsilon$
		\STATE Define: $\mathcal{F}(\boldsymbol{\tau}_N)=\!\!\!\tau_i\log_2\!\Big(\!1\!+b_i\!\frac{\!\tau_0+\sum_{j=1}^{i-1}\! \tau_j\!}{\tau_i}\!\Big)\!$\\ 		    $\,\,\,\,\,\,\,\,\,\,\,\,\,\,\,\,\,\,\,\,\,\,g(\boldsymbol{\tau}_N)\!=\!\!P_{c_D}\big(\!\tau_0\!+\!\sum_{i=1}^{N-1}\tau_i\big)+P_{c_U}\sum_{i=1}^N\tau_i$
		\STATE $n\leftarrow 0$
		\STATE Initialize $\lambda$, $\kappa_1=\kappa_2=\cdots=\kappa_N=0$
		\REPEAT
		\STATE $r\leftarrow 0$
			\REPEAT	
			\FOR {$i=1$ to $N$}
				\IF{$i\neq N$}
					\STATE{Determine $\phi_i$ using (\ref{Eq.phi_i})}
				\ELSE
					\STATE{Determine $\phi_N$ using (\ref{Eq.phi_N})}				
				\ENDIF
				\STATE{Compute $z_i$ using (\ref{eq.zstr})}
			\ENDFOR	
			\FOR{$i=N$ to $1$}	
				\IF{$i=N$}
					\STATE{Update $\tau_N$ using (\ref{eq.taustrN})}
				\ELSE
					\STATE{Update $\tau_i$ using (\ref{eq.taustri})}
				\ENDIF
			\ENDFOR				 	
			\STATE{Update $\tau_0=1-\sum_{i=1}^N\tau_i$}
			\STATE{$r\leftarrow r+1$}
			\FOR { $i=1$  to $N$}
			\IF {$\mathcal{R}_i-\mathcal{R}_{min}^i <\epsilon$}
				\STATE{$\kappa_i\neq 0$}
				\STATE{Update $\kappa_i$ using gradient method}
			\ENDIF
			\ENDFOR
			 \UNTIL{$\mathcal{R}_i-\mathcal{R}_{min}^i>\epsilon$}		
		\STATE {Determine $\Delta_n=\mathcal{F}(\boldsymbol{\tau}_N)-\lambda_n g(\boldsymbol{\tau}_N)$}
		\STATE {$\lambda_{n+1}=\frac{\mathcal{F}(\boldsymbol{\tau}_N)}{g(\boldsymbol{\tau}_N)}$}
		\STATE {$n\leftarrow n+1$}		
		\UNTIL {$|\Delta_n|<\epsilon$}	
		\STATE Set $\tau_0^*=\tau_0$ and $\tau_i^*=\tau_i$.
	\end{algorithmic}
\end{algorithm}
\section{Impact of statistical QoS constraints}
In this section, we analyze the impact of QoS constraints on the optimal time allocation strategies that target the maximization of the system energy efficiency. Since effective capacity describes the maximum constant data arrival rates, i.e. characterizes the throughput in the presence of delay-limited data sources, we focus on the effective-EE to determine the number arriving bits that can be supported per one joule of consumed energy by the system in the presence of statistical queuing constraints.

Let us first address half-duplex operation. Since UEs harvest energy simultaneously and send information-bearing signals to the AP using NOMA, harvesting time becomes the only parameter to optimize for maximum performance. In the case in which each user harvests energy to support data transfer with half-duplex operation, the corresponding effective capacity expression of user $i$ given in (\ref{eq.effC}) is modified by incorporating the additional parameter $\tau_0$, i.e., the harvesting interval, as follows:\\
\begin{equation}
\label{eq.effCi}
\begin{split}
C_i^e(\theta_i,\tau_0)&=-\frac{1}{T\theta_i}\log \left(\!\E\Bigg\{e^{-(1-\tau_0)\theta_i\log_2\Big(1+\frac{\alpha_i\tau_0}{1+\omega_i\tau_0}\Big)}\Bigg\}\right).
\end{split}
\end{equation}
The sum effective capacity of users transmitting through a multiple access channel can be determined by summing up the individual effective capacities:
\begin{equation}
\label{eq.effce}
C^e(\boldsymbol{\theta},\tau_0)=\sum_{i=1}^N C_i^e(\theta_i,\tau_0)
\end{equation}
where the vector of QoS exponents of different users is denoted as $\boldsymbol{\theta}=[\theta_1,\theta_2,\cdots,\theta_N]$.
Now, the optimization problem for maximizing the effective-EE with half-duplex operation is formulated as follows:
\begin{subequations}
	\begin{align}
	(\text{PR:4a})\,\,\,\max_{\tau_0}\,\,\,& \frac{-\mathlarger{\mathlarger{\sum}}_{i=1}^N\log\left(\E\Bigg\{e^{-(1-\tau_0)\theta_i\log_2\Big(1+\frac{\alpha_i\tau_0}{1+\omega_i\tau_0}\Big)}\Bigg\}\!\right)}{T\theta_i\E\Big\{\tau_0P_{DT}+P_{c_U}(1-\tau_0)\Big\}}\\
	\label{eq.const2aa}
	\text{subject to}\,\,\,&\tau_0(1- \tau_0)\leq 0
	\end{align}
\label{eq:effective-EE-max-halfduplex}
\end{subequations}
Note that in (\ref{eq:effective-EE-max-halfduplex}), the objective function is the system effective energy efficiency while the constraint specifies the feasible range of the harvesting interval. Note further that effective-EE above is defined as a long-term averaged energy efficiency metric due to the presence of expectations.
\begin{Lem} \label{lemma:lemma3}
The effective-EE of energy-harvesting UEs with half-duplex protocol is pseudo-concave with respect to the harvesting interval $\tau_0$.
\end{Lem}

\emph{Proof}: See Appendix \ref{app:lemma3proof}.

Based on Lemma 3, the objective function of (PR:4a) is pseudo-concave and hence the problem is a concave-linear fractional problem, and the optimization procedure described in Section II can easily be applied to obtain the optimal solution.
\begin{figure*}
	\centering
	\begin{subfigure}[b]{0.32\textwidth}
		 \includegraphics[width=\textwidth]{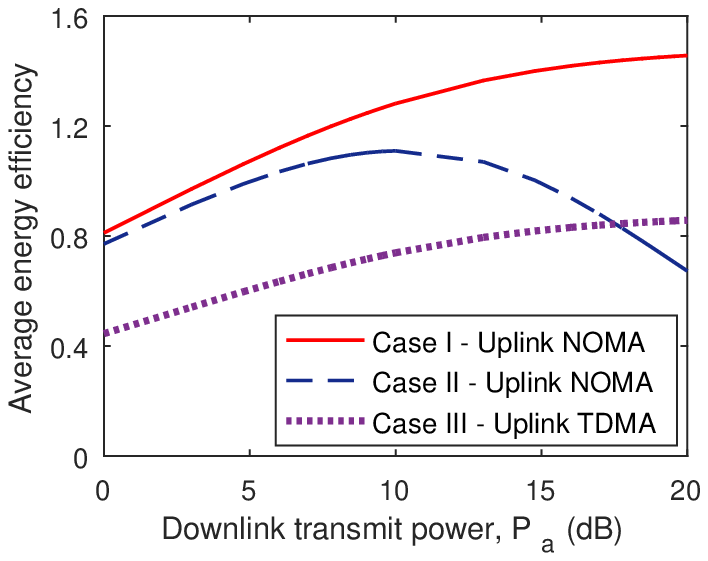}
		\caption{Average energy efficiency $\eta_{HD}$ vs. $P_a$}
		\label{Fig.EE-Pa}
	\end{subfigure}
	~ 
	\centering
	\begin{subfigure}[b]{0.32\textwidth}
		 \includegraphics[width=\textwidth]{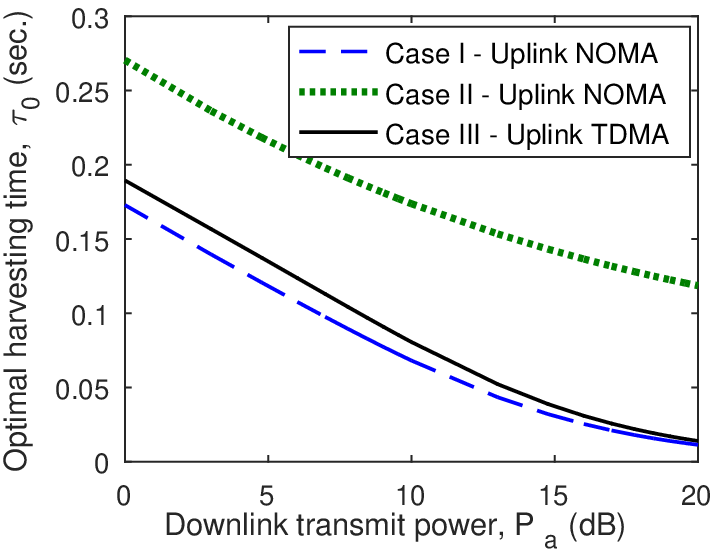}
		\caption{Optimal harvesting time vs. $P_a$}
		\label{Fig.tau0-Pa}
	\end{subfigure}
	\begin{subfigure}[b]{0.32\textwidth}
		 \includegraphics[width=\textwidth]{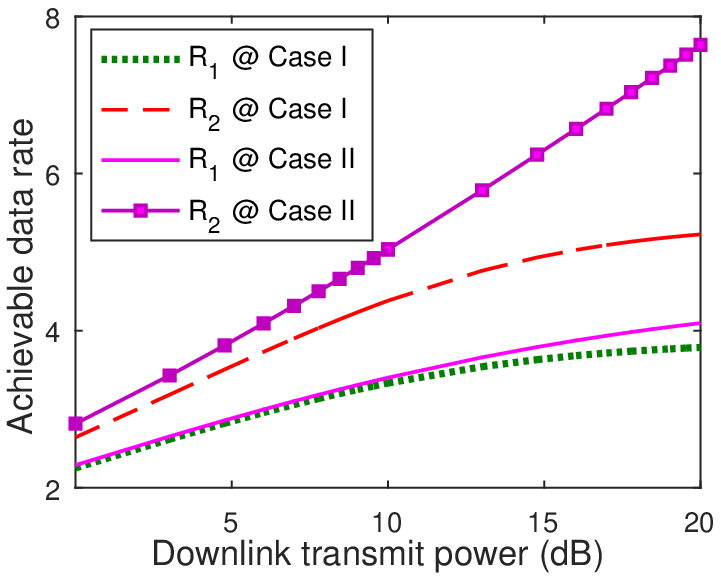}
		\caption{Achievable rates vs. $P_a$}
		\label{Fig.R-Pa}
	\end{subfigure}
	\caption{Impact of downlink transmit power level $P_a$ with half-duplex downlink-uplink operation}	
	\label{Fig.HD}
	 $\noindent\makebox[\linewidth]{\rule{18.5cm}{0.5pt}}$
\end{figure*}
Similarly, for the asynchronous transmission scenario, we have
\begin{subequations}
	\begin{align}
           (\text{PR:4b})\,\,\,\max_{\tau_0,\boldsymbol{\tau}_N}\,\,\,&\frac{\sum_{i=1}^N C_i^e(\theta_i,\tau_N)}{\E\Big\{P_{DT}\Big(\tau_0+\sum_{i=1}^{N-1}\tau_i\Big)+P_{c_U}\sum_{i=1}^N\tau_i\Big\}}\\
	\label{eq.const2aa}
	\text{subject to}\,\,\,&\sum_{i=1}^N\tau_i\leq 1- \tau_0
	\end{align}
\end{subequations}
where
\begin{equation}
\label{eq.effCi}
\begin{split}
C_i^e(\theta_i,\boldsymbol{\tau}_N)&=-\frac{1}{T\theta_i}\log\Bigg(\E\Bigg\{e^{-\Phi_i\log_2\left(1+\frac{b_ip_i}{1+\sum_{l=1}^{i-1}b_lp_l}\right)}\Bigg\}\Bigg)
\end{split}
\end{equation}
with $\boldsymbol{\theta}=[\theta_1,\theta_2,\cdots,\theta_N]$, $\Phi_i=\tau_i\theta_i$ and $p_i=\frac{\sum_{k=0}^{i-1}\tau_k}{\sum_{j=i}^N \tau_j}$. Again, similar algorithmic approaches as in Section II can be employed to solve (PR:4b).
\begin{figure*}
	\centering
	\begin{subfigure}[b]{0.32\textwidth}
		 \includegraphics[width=\textwidth]{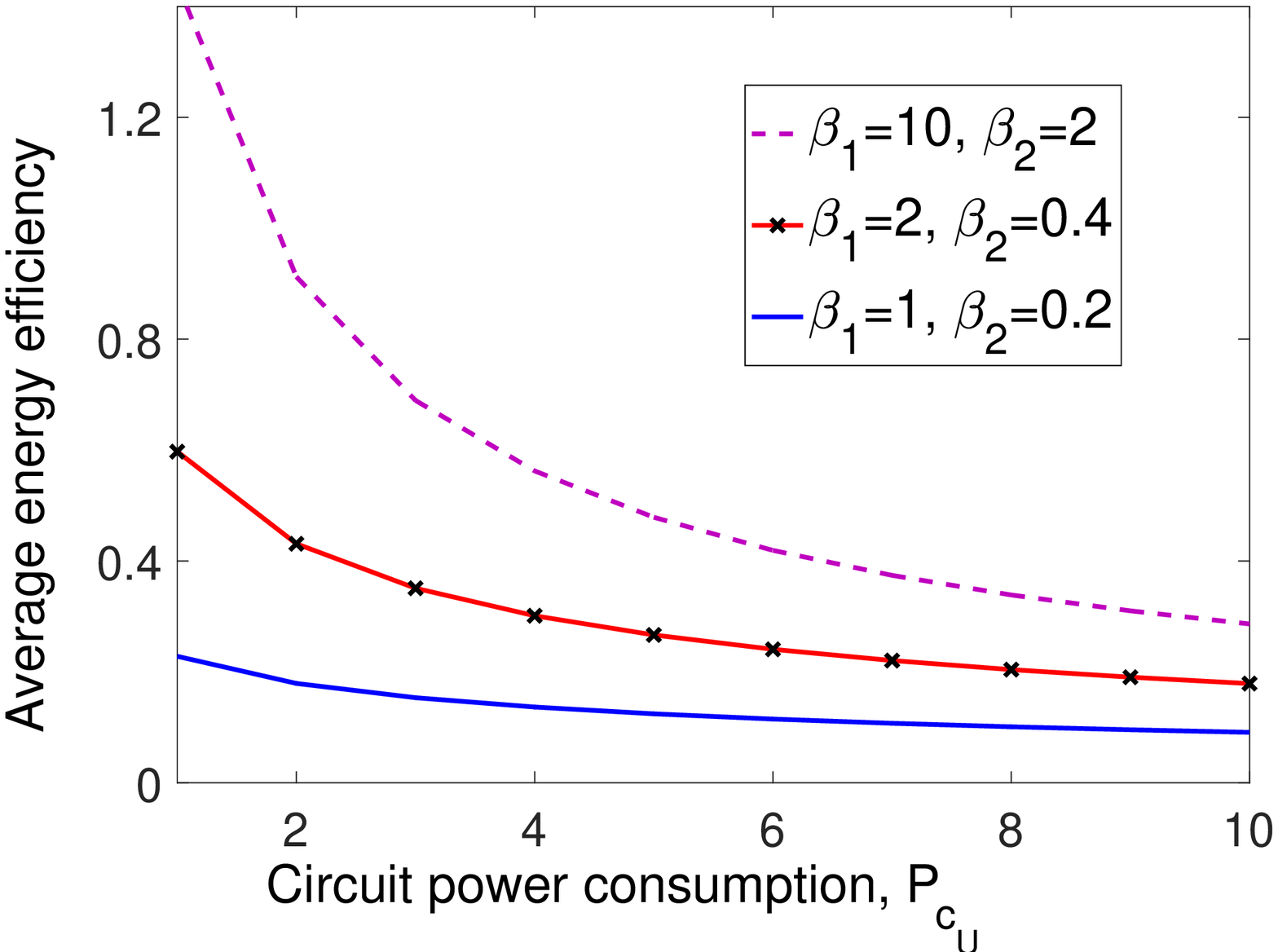}
		\caption{Average energy efficiency vs. $P_{c_U}$}
		\label{Fig.EE-Pc_AT}
	\end{subfigure}
	~ 
	\centering
	\begin{subfigure}[b]{0.32\textwidth}
		 \includegraphics[width=\textwidth]{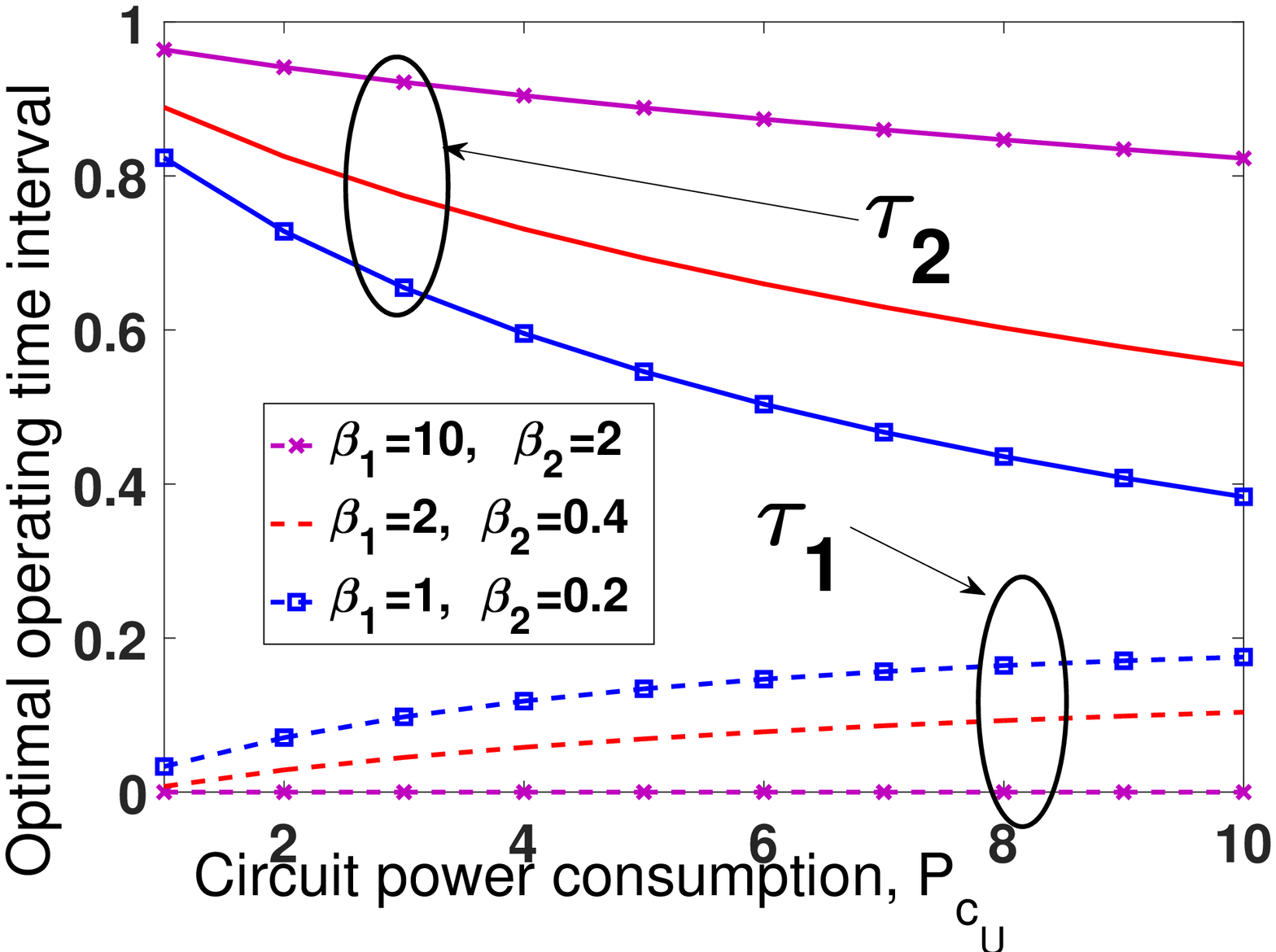}
		\caption{Operating intervals vs. $P_{c_U}$}
		\label{Fig.tau-Pc_AT}
	\end{subfigure}
	\begin{subfigure}[b]{0.32\textwidth}
		 \includegraphics[width=\textwidth]{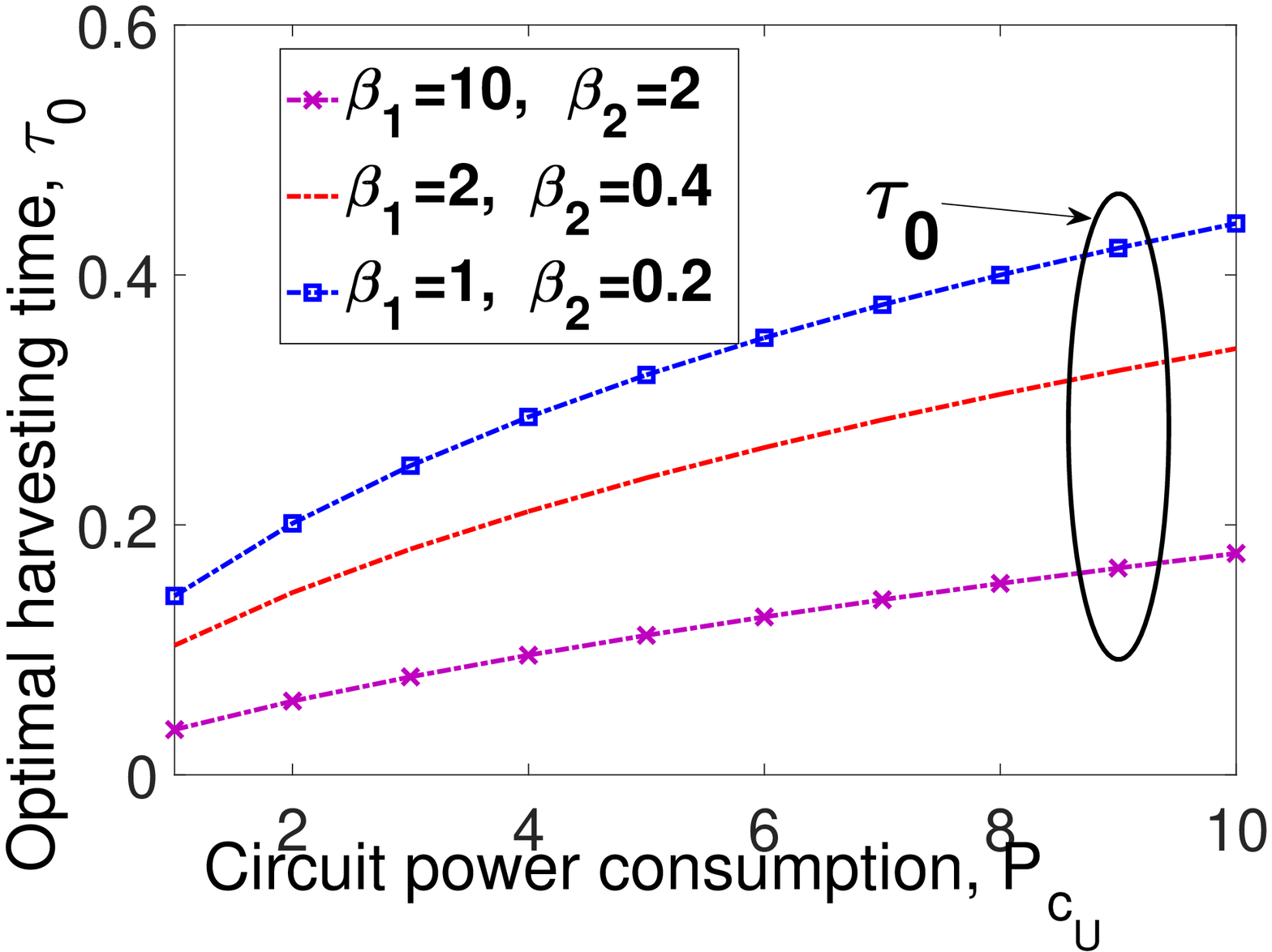}
		\caption{Harvesting time vs. $P_{c_U}$}
		\label{Fig.tau0-Pc_AT}
 	\end{subfigure}
	\caption{Effect of uplink (receiver) circuit power consumption $P_{c_U}$ on the performance with asynchronous transmission}	
	\label{Fig.AT-Pc}
	 $\noindent\makebox[\linewidth]{\rule{18cm}{0.5pt}}$
\end{figure*}
\begin{figure}[!h]
	\centering
	 \includegraphics[width=0.35\textwidth]{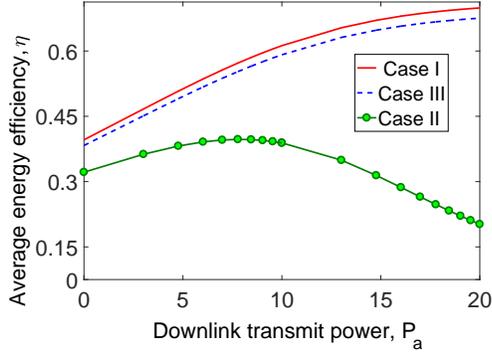}
	\caption{Performance gain of asynchronous transmission}
	\label{Fig.EE-Pa_AT}
	 $\noindent\makebox[\linewidth]{\rule{9cm}{0.5pt}}$
\end{figure}
\begin{figure*}
	\centering
	\begin{subfigure}[b]{0.4\textwidth}
		 \includegraphics[width=\textwidth]{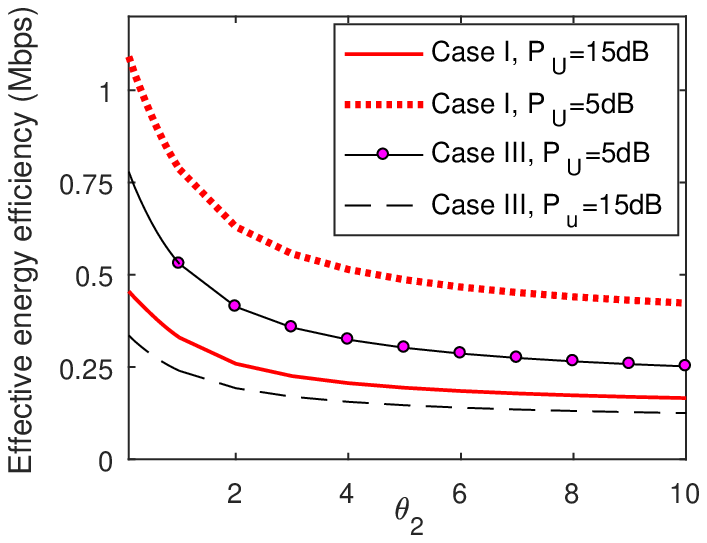}
		\caption{Effective EE vs. user 2 QoS exponent $\theta_2$}
		\label{Fig.EE-theta2}
	\end{subfigure}
	~ 
	\centering
	\begin{subfigure}[b]{0.4\textwidth}
		 \includegraphics[width=\textwidth]{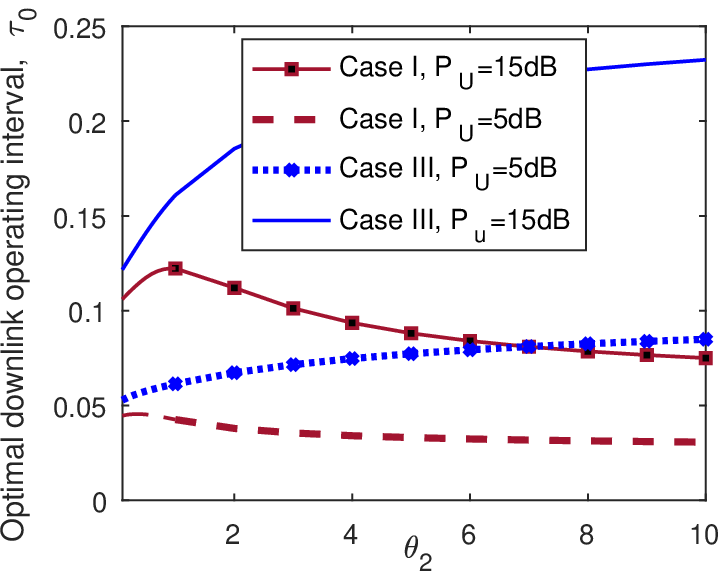}
		\caption{Downlink operating interval $\tau_0$ vs. $\theta_2$}
		\label{Fig.tau0-theta2}
	\end{subfigure}
	\caption{Impact of QoS parameter $\theta$ and circuit power consumption on the performance characteristics}	
	\label{Fig.HD-theta}
	 $\noindent\makebox[\linewidth]{\rule{18cm}{0.5pt}}$
\end{figure*}
\section{Numerical Analysis}
In this section, we provide numerical results considering two energy harvesting UEs communicating with an AP. We assume that the channel gain for the link between UE $i\in\{1,2\}$ and the AP is exponentially distributed with mean $\beta_i$. In order to compare the performance gains, we consider three cases denoted by I, II and III. In the first case, we focus on energy efficient solutions that are obtained for both half-duplex and asynchronous transmissions using uplink NOMA, as discussed in this paper. In the second case, we determine the throughput maximizing time allocations for the same problems, and in the last case we apply energy-efficiency maximization for time-division multiple access (TDMA). Additionally, we consider two values, i.e., $P_{c_U}=5dB$ and $P_{c_U}=15dB$, for the uplink power consumption in order to capture its impact on the overall characteristics.

\indent Fig. \ref{Fig.HD} illustrates the performance of WPCN operating in half duplex mode with uplink NOMA. According to Fig. \ref{Fig.EE-Pa}, we observe that broadcasting the downlink signal at a higher power level improves the system energy efficiency for case I. This is because, as $P_a$ increases, more energy can be harvested over a smaller time duration $\tau_0$ as shown in Fig. \ref{Fig.tau0-Pa}, and hence the UEs get an opportunity to transfer information over a longer time period which in turn benefits the energy efficiency. Meanwhile, comparing case I and case II as shown in Fig. \ref{Fig.EE-Pa}, we notice that allocating the harvesting interval with the goal to maximize the throughput hurts the system energy efficiency, and the degradation becomes more significant at higher values of the downlink transmit power level. Furthermore, comparing all the three cases, we observe that NOMA based uplink information transfer outperforms uplink TDMA. Intuitively, EE optimized system will be more energy-efficient than throughput optimizing systems, and hence it is expected that case I achieves better performance compared with case II. Meanwhile, for lower downlink transmit power, throughput maximizing time allocation strategy using uplink NOMA is more energy-efficient than the energy-efficiency maximizing policy for uplink TDMA as shown in Fig. 3a. One reason for this could be that the latter approach requires more time for downlink operation, and this means more energy consumption over the interval $\tau_0$. However, this is not necessarily the case for very high $P_a$ values. 
In regard to the achievable data rates, intuitively we expect throughput in case II to be higher than that in case I, and Fig. \ref{Fig.R-Pa} demonstrates this fact, i.e., $R_{i} @ \text{ Case II}$ $>$ $R_{i} @ \text{ Case I}$ $\forall P_a$ where $i\in\{1,2\}$. As can be seen from the figure, the performance gain in terms of throughput is not significant at lower downlink transmit power levels, but this changes as $P_a$ increases. Furthermore, comparing the individual data rates of UE 1 and UE 2, the latter transmits data to the AP at higher rates in both cases. However, the gap between $R_1$ and $R_2$ is smaller in case I than in case II, and this reveals that energy-efficient strategy of NOMA scheme encourages fairness in data rates among UEs. It is also interesting to observe that the performance difference between uplink NOMA and uplink TDMA lies in the optimal time allocated to each UE to transmit data uplink to the receiver. As can be seen from Fig. \ref{Fig.tau0-Pa}, energy-efficient downlink operating intervals for Case I and Case III are very close specially for higher $P_a$ values, and hence the way uplink interval is allocated determines the system performance.

Figs. \ref{Fig.AT-Pc} and \ref{Fig.EE-Pa_AT} demonstrate the system performance and the corresponding optimal operating parameters when the UEs operate in the asynchronous transmission mode. As can be seen in Fig. \ref{Fig.EE-Pc_AT}, the system energy efficiency decreases with an increase in circuit power consumption at the receiving end, and this tradeoff characteristic depends on the wireless link power gain between each UE and the receiver. More specifically, when UEs have relatively favorable channel conditions, i.e., higher gains, each incremental circuit power hurts the EE significantly. This is because the energy efficient strategy dictates both users to harvest and to transmit synchronously, i.e, $\tau_1=0$ (as seen in the case with $\beta_1 = 10$ and $\beta_2 = 2$). Besides, under this channel condition, we observe that $\tau_2>\tau_0$ while $\tau_1=0$ and this implies more time is allocated to information transfer than energy harvesting. On the other hand, worse channel characteristics lead to $\tau_1\neq 0$, and the reduction in the average energy efficiency decreases as AP circuit power consumption increases. In such a case, more time is allocated to energy harvesting, and this in turn reduces throughput and system energy efficiency. Intuitively, the impact of WPT circuit power depends on the downlink transmit power level and for higher values of $P_a$, i.e., when $P_a\gg P_{c_D}$, the change in EE along with $P_{c_D}$ is expected to be small. Meanwhile, from Fig. \ref{Fig.EE-Pa_AT}, we observe that EE increases with downlink transmit power level (similarly as discussed earlier for the half-duplex operation) unless throughput maximization is the goal as in Case II. In addition, WPCN with uplink NOMA achieves better energy efficiency compared to case III in which TDMA is considered.

\indent The impact of QoS parameter on the optimal time allocation strategy and the corresponding system energy efficiency is illustrated in Figs. \ref{Fig.EE-theta2} and \ref{Fig.tau0-theta2}. In general, stricter QoS constraint (i.e., higher value for the QoS exponent $\theta$) degrades the system energy efficiency as can be seen from Fig. \ref{Fig.EE-theta2}, and higher circuit power consumption hurts the efficiency further as expected. In addition, we observe that uplink NOMA outperforms the TDMA approach regardless of the value of $\theta$. However, the performance gain due to uplink NOMA diminishes with an increase in the aggregate circuit power consumption of UEs. In regard to the optimal time allocation strategy for TDMA, we observe that higher $\theta$ forces to allocate more time for energy harvesting, i.e., leads to increased $\tau_0$, which in turn reduces the time for uplink information transfer.

\section{Conclusion}
In this paper, we have considered energy efficiency as a performance metric, and we have investigated impact of uplink-NOMA on the overall performance of energy-harvesting communication networks. We have taken into account half-duplex and asynchronous transmission downlink-uplink operation modes, and formulated optimization problems in both cases focusing on the maximization of the system energy efficiency. Since these are concave-linear fractional programming problems, Dinkelbach's method can be directly applied. With this, we have obtained closed-form characterizations for the optimal time itervals and provided an algorithm to obtain the optimal solution for half duplex operation. Meanwhile, because of the difficulty in obtaining closed-form solutions for asynchronous transmission, we have analyzed the optimal solution using standard numerical tools. Finally, several insightful observations have been made through numerical results. According to the these results, we have seen that downlink transmit power improves the system energy efficiency. In addition, circuit power consumption hurts EE, but this depends on the channel characteristics. Time intervals for energy harvesting and data transmission display intricate dependence on system parameters and operational modes. Finally, we have noted that stricter delay constraints lead to degradation in energy efficiency.

\appendix

\subsection{Proof of Lemma \ref{lemma:lemma2}} \label{app:lemma2proof}
For the case of two users, we have $\tau_1$ and $\tau_2$, and hence we show that the throughput in each interval is jointly concave with respect the operating intervals:\\
(i) During $\tau_1$: In this case, only UE 1 transmits information uplink to the access point, i.e.,
\begin{equation}
R_{sum}^1=\tau_1\log_2\left(a_1+\frac{b_1}{\tau_1+\tau_2}\right).
\end{equation}
Let $\bold{F}^i$ denote the Hessian matrix of $R_{sum}^i$ with respect to $\tau_1$ and $\tau_2$. Then, applying the second-order derivatives, we get
\begin{equation}
\begin{split}
F_{11}^1=\frac{\partial^2 R_{sum}^1}{\partial \tau_1^2}&=-\frac{b_1(2a_1\tau_2(\tau_2+\tau_1))+b_1(2\tau_2+\tau_1)}{(\tau_1+\tau_2)^2(a_1(\tau_1+\tau_2)+b_1)^2}\\
F_{22}^1=\frac{\partial^2 R_{sum}^1}{\partial \tau_2^2}&=\frac{b_1\tau_1(2a_1(\tau_2+\tau_1))+b_1}{(\tau_1+\tau_2)^2(a_1(\tau_1+\tau_2)+b_1)^2}\\
F_{12}^1=\frac{\partial^2 R_{sum}^1}{\partial \tau_1\partial\tau_2}&=-\frac{b_1(a_1(\tau_2-\tau_1)(\tau_2+\tau_1)+b_1\tau_2)}{(\tau_1+\tau_2)^2(a_1(\tau_1+\tau_2)+b_1)^2}.
\end{split}
\end{equation}
Since $F_{11}^1F_{22}^1-(F_{12}^1)^2\leq0$, the Hessian $\bold{F}$ is a negative semi-definite matrix. Based on Theorem 21.5 given in \cite{Simon}, $R_{sum}^1$ is a jointly concave function of $\tau_1$ and $\tau_2$.\\
(ii) During $\tau_2$: Here, both UE 1 and UE 2 are transmitting, and the corresponding achievable sum-rate capacity is given as
\begin{equation}
R_{sum}^2=\tau_2\log_2\left(a_2+\frac{b_1}{\tau_1+\tau_2}+\frac{b_2}{\tau_2}\right),
\end{equation} and applying the second order derivative, we get
\begin{equation}
\begin{split}
F_{11}^2=\frac{\partial^2 R_{sum}^2}{\partial \tau_1^2}&=\frac{2b_1(\tau_1+\tau_2)(a_2+\frac{b_2}{\tau_2})+b_1^2}{(\tau_1+\tau_2)^4(a_2+\frac{b_1}{\tau_1+\tau_2}+\frac{b_2}{\tau_2})^2}\\
F_{22}^2=\frac{\partial^2 R_{sum}^2}{\partial \tau_2^2}&=-\frac{\frac{2b_1y}{(\tau_1+\tau_2)^3}}{a_2+\frac{b_1}{\tau_1+\tau_2}+\frac{b_2}{\tau_2}}-\frac{\tau_1(-\frac{b_2}{\tau_2^2}-\frac{b_1}{(\tau_1+\tau_2)^2})^2}{(a_2+\frac{b_1}{\tau_1+\tau_2}+\frac{b_2}{\tau_2})^2}\\
F_{12}^2=\frac{\partial^2 R_{sum}^2}{\partial \tau_1\partial\tau_2}&=-\frac{b_1\tau_2(a_2\tau_2(\tau_1-\tau_2)+\tau_1(2b_2+b_1\frac{\tau_2}{\tau_1+\tau_2}))}{(\tau_1+\tau_2)^3(\tau_2(a_2+\frac{b_1}{\tau_1+\tau_2})+b_2)^2}.
\end{split}
\end{equation}
\begin{figure*}
	\begin{equation}
	\label{eq.2nd}
	\begin{split}
	\ln(2)\frac{\partial h(\tau_B)}{\partial \tau_B}&=-\log\bigg(1+\frac{a_i\tau_B}{(1-\tau_B)+\tau_B a_\ast}\bigg)+(1-\tau_B)\frac{a_i}{\Big(1+\tau_B(a_i+a_\ast-1)\Big)\Big(1-\tau_B+a_\ast\tau_B\Big)}\\
	\ln(2)\frac{\partial^2 h(\tau_B)}{\partial \tau_B^2}
    &=-\left[\frac{\Big(1-\tau_B\Big)\Big(2a_ia_\ast+a_i^2\Big)+2a_ia_\ast\tau_B\Big(a_\ast+a_i\Big)}{\Big(1-\tau_B+a_\ast\tau_B\Big)^2\Big(1-\tau_B+a_i\tau_B+a_\ast\tau_B\Big)^2}\right]
	\end{split}
	\end{equation}
$\noindent\makebox[\linewidth]{\rule{19cm}{1pt}}$
\end{figure*}
It is obvious that $F_{11}^2F_{22}^2-(F_{12}^2)^2\leq 0$ and hence $R_{sum}^2$ is also a jointly concave function of the operating intervals using a similar argument as stated above. This completes the proof.\\~\\
\subsection{Proof of Lemma \ref{lemma:lemma3}}\label{app:lemma3proof}
\textit{Proof}: First, let $h_i(\tau_B)=(1-\tau_B)\log_2\Big(1+\frac{a_i\tau_B}{1+\big(a_\ast-1\big)\tau_B}\Big)$ and $H_i(\tau_B)=-\frac{1}{\theta_i}\log\big(\E\{e^{-\theta_ih_i(\tau_B)}\}\big)$ where $\theta_i$'s are given.
Applying the second-order derivative criterion to $h_i(\tau_B)$, it can be inferred from (\ref{eq.2nd}) given at the top of
the next page that $\frac{\partial^2 h_i(\tau_B)}{\partial \tau_B^2}<0$ for all $\tau_B\in(0,1)$, and hence $h_i(\tau_B)$ is concave or $-h_i(\tau_B)$ is convex in the domain set. This implies that $e^{-h(\tau_B)}$ is log-convex, and $\E\{e^{-h(\tau_B)}\}$ is log-convex as well since log-convexity is preserved under sums \cite{boyd}. Noting that $\log(g(\cdot))$ is convex for log-convex $g(\cdot)$, clearly $H(\tau_B)$ is a concave function of $\tau_B$ for $0<\tau_B<1$.
Meanwhile, the sum effective capacity can be re-written as
\begin{equation}
C^e(\tau_B)=\frac{H_{sm}(\tau_B)}{T}
\end{equation}
where $H_{sm}=\sum_{i=1}^N H_i(\tau_B,\theta_i)$. Since convexity/concavity is preserved under sums, it is obvious that $H_{sm}$ is also a concave function. Thus, $C^e(\tau_B)$, is a concave function, completing the proof.
\end{spacing}
\begin{spacing}{1.5}

\end{spacing}

\begin{thebibliography}{99}
\bibitem{LuX} X. Lu, P. Wang, D. Niyato, D. I. Kim, Z. Han, ``Wireless charging technologies: Fundamentals, standards, and network applications," \textit{IEEE Commun. Surveys \& Tutorials}, vol. 18, no. 2, pp. 1413--1452, 2016.
\bibitem{XieL} L. Xie, Y. Shi, Y. T. Hou, and A. Lou, ``Wireless power transfer and applications to sensor networks,” \textit{IEEE Wireless Commun.}, vol. 20, no. 4, pp. 140--145, Aug. 2013.
\bibitem{Ulu1} S.Ulukus, A. Yener, E. Erkip, O. Simeone, M. Zorzi, P. Grover, and K. Huang, ``Energy harvesting wireless communications: A review of recent advances,�  \textit{IEEE J. Sel. Areas Commun.}, vol. 33, no. 3, pp. 360--381, Mar. 2015.
\bibitem{ZhongC} C. Zhong, X. Chen, Z. Zhang, G. K. Karagiannidis, ``Wireless-powered communications: performance analysis and optimization," \textit{IEEE Trans. Commun.}, vol. 63, no. 12, pp. 1-1, Dec. 2015.
\bibitem{Hyungisk} H. Ju, R. Zhang, ``Throughput maximization in wireless powered communication networks," \textit{IEEE Trans. on Wireless Commun.}, vol. 13, no. 1, pp. 418--428, Jan, 2014.
\bibitem{Zhaof} F. Zhao, L. Wei, and H. Chen, ``Optimal time allocation for wireless information and power transfer in wireless powered communication systems," DOI: 10.1109/TVT.2015.2416272
\bibitem{Hwang} D. Hwang, D. I. Kim, and T. J. Lee, ``Throughput maximization for multiuser MIMO wireless powered communication networks," DOI: 10.1109/TVT.2015.2453206.
\bibitem{Leeh} H. Lee, K. J. Lee, H. B. Kong, and I. Lee, ``Sum rate maximization for multi-user MIMO wireless powered communication networks,"  DOI: 10.1109/TVT.2016.2515607.
\bibitem{Ju} H. Ju, and R. Zhang, ``Optimal resource allocation in full-duplex wireless-powered communication network," \textit{IEEE Trans. Commun.}, vol. 62, no. 10, pp. 3528--3540, Oct. 2014.
\bibitem{Kang} X. Kang, C. K. Ho, and S. Sun, ``Full-duplex wireless-powered communication network with energy causality," \textit{IEEE Trans. Wireless Commun.}, vo. 14, no. 10, pp. 5539--5551, Oct. 2015.
\bibitem{Teddy} T. A. Zewde, and M. C. Gursoy, ``Wireless-powered communication under statistical quality of service constraints," \textit{Proc. of IEEE International Conference on Commun. (ICC)}, Kuala Lumpur, Malaysia, 2016.
\bibitem{Dai} L. Dai, B. Wang, Y. Yuan, S. Han, C.L I, and Z. Wang, ``Non-orthogonal multiple access for 5G: Solutions, challenges, opportunities, and future research trends," \textit{IEEE Commun. Mag.}, vol. 53, no. 9, pp. 74--91, Sep. 2015.
\bibitem{Islam} S. R. Islam, N. Avazov, O. A. Dobre, and K.S. Kwak, ``Power-domain non-orthogonal multiple access (NOMA) in 5G systems: potentials and challenges," https://arxiv.org/abs/1609.06261, DOI: 10.1109/COMST.2016.2621116,
\bibitem{Datta} S. N. Datta, and S. Kalyanasundaram, ``Optimal power allocation and user selection in non-orthogonal multiple access systems," \textit{Proc. of IEEE WCNC}, Apr. 2016.
\bibitem{Lei} L. Lei, D. Yuan, C. K. Ho, and S. Sun, ``Power and channel allocation for non-orthogonal multiple access in 5G systems: tractability and computation," \textit{IEEE Trans. Commun.}, pp. 8580 - 8594, vol. 15, no. 12, Dec. 2016.
\bibitem{Han} W. Han, Y. Zhang, X. Wang, J. Li, M. Sheng, and X. Ma, ``Orthogonal power division multiple access: A green communicaiton prespective," \textit{IEEE Jour. Sel. Area Commun.}, vol. 34, no. 12, pp. 3828--3842, Dec. 2016.
\bibitem{BuzziS} S. Buzzi, C. L. I, T. E. Klein, H. V. Poor, C. Yang, and A. Zappone, ``Survey of energy-efficient techniques for 5G networks and challenges ahead," \textit{IEEE J. Sel. Areas Commun.}, vol. 34, no. 4, pp. 697--709, 2016.
\bibitem{Huu} Q. T. Vien, T. A. Le, B. Barn, C. V. Phan, ``Optimising energy efficiency of non-orthogonal multiple access for wireless backhaul in heterogenous cloud radio access network," \textit{IET Commun.}, 2016.
\bibitem{Huu2} H. Q. Tran, P. Q. Truong, C. V. Phan, Q. T. Vien, ``On the energy efficiency of NOMA for wireless backhaul in multi-tier heterogenous CRAN," \textit{Proc. of Intern. Conf. Recent Advances in SigTelCom}, 2017.
\bibitem{FF} F. Fang, H. Zhang, J. Cheng, V. C.M. Leung,``Energy-efficient resource allocation for downlink non-orthogonal multiple access network," \textit{IEEE Tran. Commun.}, vol. 64, no. 9, pp. 3722-3732, Sep. 2016.
\bibitem{SunQ} Q. Sun, S. Han, C.L. I, Z. Pan, ``Energy efficeincy optimization for fading MIMO non-orthogonal multiple access system,"\textit{Proc. IEEE ICC}, 2015.
\bibitem{Zhang} Y. Zhang, H. M. Wang, T. X. Zheng, and Q. Yang, ``Energy-efficient transmission design in non-orthogonal multiple access," \textit{IEEE Trans. Vehicular Technology}, Jun. 2016.
\bibitem{Diam} P. D. Diamantoulakis, K. N. Pappi, Z. Ding, and G. K. Karagiannidis, ``Wireless powered communications with non-orthogonal multiple access," \textit{IEEE Trans. Wireless Commun.}, pp. 8422 - 8436, vol. 15, no. 12, Dec. 2016.
\bibitem{Chin} H. Chingoska, Z. Hadzi-Velkov, I. Nikoloska, and N. Zlatanov, ``Resource allocation in wireless powered communication networks with non-orthogonal multiple access," \textit{IEEE Wireless Commun. Letts.}, pp. 684 - 687, vo. 5, no. 6, Dec. 2016.
\bibitem{Yuan} Y. Yaun, Z. Ding, ``The application of non-orthogonal multiple access in wireless powered communication networks," \textit{Proc. IEEE SPAWC}, pp. 1-5, Jul. 2016.
\bibitem{Zding} Z. Ding, X. Lei, and G. K. Karagiannidis,  ``A Survey on Non-Orthogonal Multiple Access for 5G Networks: Research Challenges and Future Trends,”  \textit{IEEE J. Sel. Areas Commun.}, pp.  2181 - 2195, vol. 35, no. 10, 2017.
\bibitem{Lsong} L. Song, Y. Li, Z. Ding, and H. V. Poor, ``Resource Management in Non-orthogonal Multiple Access Networks for 5G and Beyond," \textit{IEEE Networks}, vol. 31, n. 4, pp. 8-14, 2017.
\bibitem{Chang} C.S. Chang  and  T. Zajic, ``Effective  bandwidths  of  departure  processes  from  queues  with  time  varying  capacities," \textit{Proc. IEEE Infocom.}, pp. 1001-1009,  1995.
\bibitem{dapeng} D. Wu and R. Negi ``Effective capacity: a wireless link model for support of quality of service,"
 \emph{IEEE Trans. Wireless Commun.}, vol.2,no. 4, pp.630-643. July
 2003.
\bibitem{Alessio} A. Zappone, and E. Jorswieck, ``Energy efficiency in wireless networks via fractional programming theory," \textit{Foundations and Trends in Communications and Information Theory}, vol. 11, no. 3-4, pp. 185-396, 2015.
\bibitem{Simon} C. P. Simon and L. Blume, \textit{Mathematics for Economists}, W. W. Norton \& Company Inc., New York, NY, 1994.
\bibitem{boyd}    S. Boyd and L. Vandenberghe, \textit{Convex Optimization}, Cambridge University Press, 2004.
\end{thebibliography}
\end{document}